\documentclass[journal=jacsat,manuscript=communication,layout=twocolumn]{achemso}
\usepackage{graphicx}
\usepackage{amssymb}
\usepackage{amsmath}
\usepackage{nomencl}
\usepackage{cleveref}
\usepackage{upgreek} 

\SectionsOn
\SectionNumbersOn

\author{Stefanos~D.~Anogiannakis} 
\affiliation{National Technical University of Athens, School of
  Chemical Engineering, Zografou Campus, GR-15780 Athens, Greece}

\author{Christos~Tzoumanekas} 
\affiliation{National Technical University of Athens, School of
  Chemical Engineering, Zografou Campus, GR-15780 Athens, Greece}
\alsoaffiliation{Dutch Polymer Institute (DPI), P.O. Box 902, 5600 AX
  Eindhoven, The Netherlands}
\email{tzoumanekas@gmail.com}

\author{Doros~N.~Theodorou}
\affiliation{National Technical University of Athens, School of
 Chemical Engineering, Zografou Campus, GR-15780 Athens, Greece}
\alsoaffiliation{Dutch Polymer Institute (DPI), P.O. Box 902, 5600 AX
  Eindhoven, The Netherlands}

\title{Microscopic Description of Entanglements in Polyethylene Networks and Melts:
       Strong, Weak, Pairwise, and Collective Attributes}

\makeatletter
\renewcommand\section{\@startsection{section}{1}{\z@}%
                                  {-3.5ex \@plus -1ex \@minus -.2ex}%
                                  {2.3ex \@plus.2ex}%
                                  {\normalfont\small\bfseries}}                                  
\makeatother

\makeatletter
\renewcommand\subsection{\@startsection{subsection}{1}{\z@}%
                                  {-3.5ex \@plus -1ex \@minus -.2ex}%
                                  {2.3ex \@plus.2ex}%
                                  {\normalfont\small\small\bfseries}}
\makeatother

\begin{document}
\begin{abstract}
\begin{figure} 
\begin{center}
  \includegraphics[clip,width=1.0\linewidth] {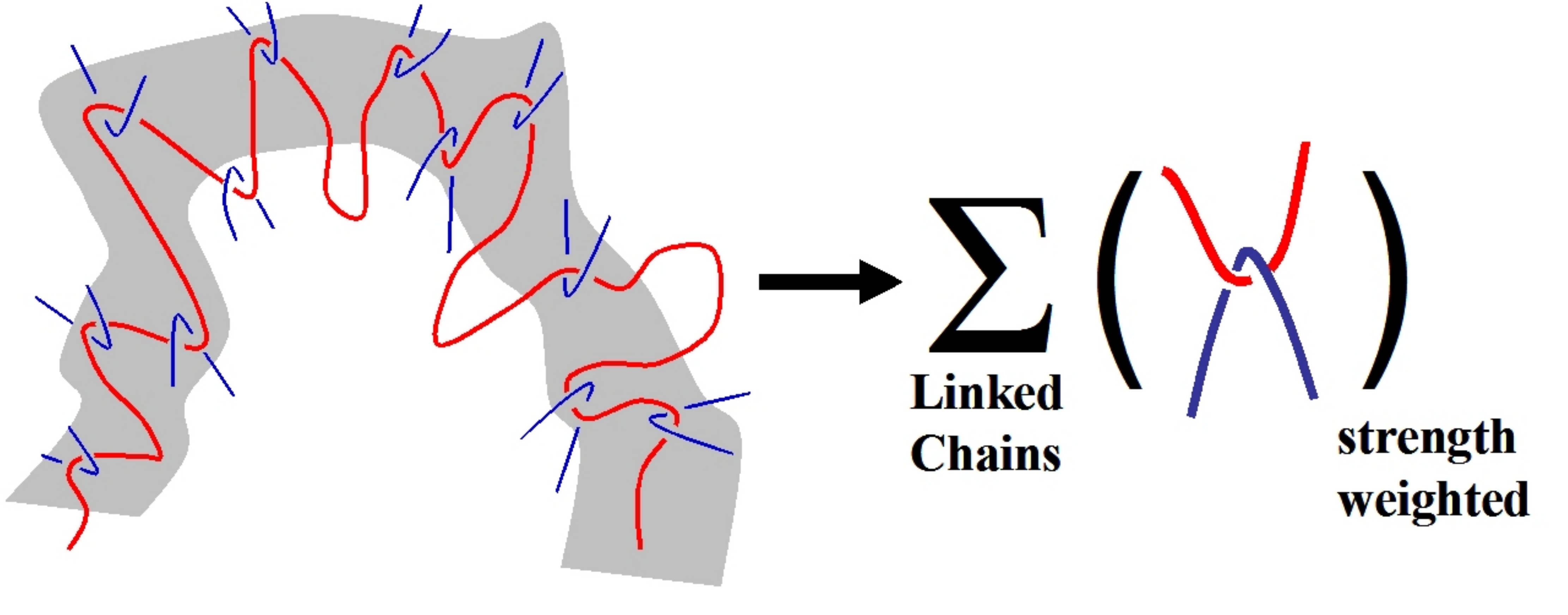}
\end{center}
\end{figure}

We present atomistic molecular dynamics simulations of two Polyethylene systems where all entanglements are 
trapped: a perfect network, and a melt with grafted chain ends. We examine microscopically
at what level topological constraints can be considered as a collective entanglement effect,
as in tube model theories, or as certain pairwise uncrossability interactions, as in slip-link models.
A {\it pairwise parameter}, which varies between these limiting cases,
shows that, for the systems studied, the character of the entanglement environment is more pairwise than 
collective. We employ a novel methodology, which analyzes entanglement constraints into a complete set of 
pairwise interactions, similar to slip links. Entanglement confinement is assembled by a plethora of links, 
with a spectrum of confinement strengths, from strong to weak. The strength of interactions
is quantified through a link `persistence', which is the fraction of time for which the links are active.
By weighting links according to their strength, we show that confinement
is imposed mainly by the strong ones, and that the weak, trapped, uncrossability interactions
cannot contribute to the low frequency modulus of an elastomer, or the plateau modulus of a melt. 
A self-consistent scheme for mapping
topological constraints to specific, strong binary links, according to a given entanglement density,
is proposed and validated. Our results demonstrate that slip links can be viewed as the strongest
pairwise interactions of a collective entanglement environment.
The methodology developed provides a basis for bridging the 
gap between atomistic simulations and mesoscopic slip link models. 
\end{abstract}
%

\section{Introduction}
\label{section:Intro}
In the molecular description of structure-property relations of polymer melts and networks, 
one of the fundamental concepts is chain entanglement. When macromolecules
interpenetrate, the term entanglements intends to describe {\it interactions}
resulting from the {\it uncrossability} of chains. In flexible polymer melts and rubbers these 
interactions alter in a universal manner their dynamical, flow, and deformation 
properties.

Molecular-based theories which aim to incorporate entanglements have first to
define them. This task poses certain difficulties. In mathematical terms,
entanglement interactions originate from topological constraints (TCs).
Thus, in a many-chain theory, and in order to derive entanglement effects 
from `first principles', a rigorous description \cite{Edwa-67b,Edwa-68,Deam-76}
invokes the use of topological invariants and certain approximations to make the 
problem tractable. In such theories, however, the range of problems that can be studied
is limited by mathematical complexity.
  
In order to deal with entangled matter, several conceptual simplifications were
introduced by Edwards \cite{Edwa-67a,Edwa-77}, de Gennes \cite{deGen-71,deGen-79}, Doi and Edwards \cite{Doi-78b,Doi-86},
and other pioneers in the field of polymer physics. The crucial simplification was the description
of entanglement through a mean field and a single chain model. In tube model theories 
\cite{Edwa-67a,deGen-79,Doi-86,Edwa-88} it is postulated that the entanglements generate a confining, 
(quadratic) mean field potential, which restricts lateral monomer fluctuations to a tube-like region 
surrounding each chain. In a perfect network, the potential attracts the monomers toward the centerline 
of the tube and chains are restricted to adopt conformations which are compatible with the tube constraint. 
In polymer melts, this confinement is not permanent, but leads to reptation \cite{deGen-71}, 
a one-dimensional diffusion of the chain along the tube.  

The axis of the tube plays the role of a coarse-grained representation of the real chain, 
and it is called the primitive path \cite{Edwa-77,Doi-78b,Doi-86} (PP). It has a random-walk like conformation, 
like the real chain, but with a larger step length, which corresponds to the entanglement molar mass
 \cite{Doi-86}, $M_{\rm e}$. The latter is determined from the plateau modulus of long chain polymer 
melts. The average tube diameter is not so well-defined; it is considered to be of the order \cite{Doi-86}
of magnitude of the PP step length; it is often assumed that the two lengths are equal. 

An alternative, discrete, localized version of the tube constraint is utilized in models employing 
slip-links \cite{Graes-77,Doi-78b,Ball-81,Edwa-86,Hua-98,Rubi-02}. 
The tube is replaced by a set of slip links along the chain, which
restrict lateral motion but permit chain sliding through them. 
The real chain is represented by the corresponding PP, which is a series of 
strands of average molar mass $M_{\rm e}$ connecting the links.
The tube model provides a basis for working out analytically the linear and nonlinear viscoelasticty 
of polymer melts \cite{Doi-86,Wat-99,McLeish-02,Rubi-03,Larson-06,Graess-08} and networks 
\cite{Gayl-87,Gayl-90,Edwa-88,Hein-88,Rubi-02,Graess-04} at a molecular level.
Slip link models \cite{Graess-04,Larson-06,Graess-08}  also allow for analytical 
\cite{Graes-77,Doi-78b,Ball-81,Edwa-86,Rubi-02,Khal-08,Schi-10} or 
stochastic simulation \cite{Hua-98,Tasa-01,Masu-01,Shan-01,Doi-03,Masu-04a,Likh-05,Nair-06,Ober-06,Masu-08,Khal-09,Schi-12} 
treatments with a molecular representation, at the single or many-chain level. 
 
The concepts conveyed by the tube and slip-link models provide the prevailing microscopic view of 
entanglement. In stochastic simulations, the slip links are usually envisaged as 
{\it binary entanglements}, a form of discrete {\it pairwise} contacts
which confine (link) the motion of two neighboring chains. 
The tube constraint, on the other hand, acts along the whole chain; {\it locally}, it is thought  
that confinement is applied {\it collectively} by many overlapping surrounding chains,
and not by specific pairwise interactions. Any gain or loss in the local pairwise interaction density
does not necessarily affect the tube diameter or length, as it does the local entanglement
density of slip-link models\cite{Larson-06} (see p 552 in ref \citenum{Graess-08}). 

A microscopic picture of entanglement is also suggested by scaling models
relating the degree of entanglement to chemical structure. Notably, the 
most successful of them are based on similar \cite{Colby-92,Heyn-00,Miln-05,Wang-07a} ideas.
A good correlation with polymer melt data is provided by the 
packing length \cite{Fett-94} model, which is based on the conjecture of Lin \cite{Lin-87}, and 
Kavassalis and Noolandi\cite{Kava-87,Kava-88}. The conjecture is in support of a collective\cite{Rubi-97,Miln-05} argument;
an entanglement results from a fixed number of polymer chains cohabiting the same volume.
On the other hand, the scaling model \cite{Colby-90,Colby-92} of Colby and Rubinstein, which is in agreement with data 
from solutions in a theta solvent, is in support of a pairwise \cite{Rubi-97,Miln-05} argument; a fixed number of binary contacts
between chains gives rise to an entanglement strand. 

Despite the success of tube and slip-link models in describing the linear (especially) and nonlinear 
rheology of polymer melts, as well as rubber elasticity, the nature of entanglements remains elusive.
It would be desirable, therefore, to probe entanglements microscopically in order to 
correct, refine and support the perspective given above with microscopic information.
Simulation is probably the method of choice, since we need access to some kind of topological information 
which is not directly accessible with spectroscopic methods. 

The question, then, is how to define entanglements in a set of densely packed chain molecules.  
A unique definition might not exist, but a certain {\it microscopic description} has to be invoked to
address the above topics. In this direction, several attempts have been made.
Assuming that entanglements lead to persistent chain interactions, 
the detection of long-lived chain contacts \cite{Gao-95,Ben-96,Wang-97,Yama-04} was the target of many simulations.
Although it was found that such contacts do exist, it is rather difficult to evolve this picture further.
Additionally, the topological origin of entanglements is overlooked.
It becomes prominent, however, in studies \cite{Ever-96,Lang-01,Orlan-04,Tzoum-11,Rosa-11,Miln-11b} employing topological
invariants. The latter can detect pairs of entangled and unentangled chains, and the `amount' of winding between the chains.
We have to note that, for open chains, true invariants do not exist. Moreover, the simplest invariant for closed chains, 
the Gauss linking number, can misclassify \cite{Edwa-68} topological constraints. In any case, the provided information 
is at the global level of a chain. Though very useful, it is not clear how to put it into the context of current tube and
slip-link model ideas. The latter require a somewhat localized description of entanglement, especially in connection with
constraint release (CR) processes. In this respect, the use of a local topological measure may be more suitable. 

Microscopic investigations of tube model ideas have greatly benefited 
from simulations which construct the PPs according to the Edwards definition \cite{Edwa-77}, i.e., by holding chain ends fixed 
and removing all chain slack from the system. The first implementation \cite{Ever-04,Suku-05} was by Everaers et al., 
but soon thereafter other methods for shrinking the chains in continuous \cite{Krog-05,Tzoum-06a}
and lattice models \cite{Shan-05,Shan-06} appeared. 
There is strong evidence \cite{Leon-05,Tzoum-06a,Tzoum-07,Kaz-07,Harm-09,Hou-10} that the PP Kuhn segment coming out from these 
methods is in very good agreement with $M_{\rm e}$ estimations from plateau modulus measurements. 
The PP representation is also popular in slip-link simulations. Thus, it is hoped that these 
coarse graining methods could serve as a bridge \cite{Tzoum-06b,Masu-10} between the different levels of description. 

The properties of PPs have been studied in several systems, such as rubbers \cite{Li-11}, melts, 
nanocomposites \cite{Rigg-09,Term-10a,Term-10b,Toepp-11,Li-12}, grafted brushes \cite{Hoy-07}, glasses \cite{Hoy-05}, 
in equilibrium or under deformation \cite{Toepp-11} and flow \cite{Baig-10c}. 
Concepts and properties such as entanglement density scaling \cite{Ever-04,Suku-05,Uch-08}, 
entanglement molecular weight \cite{Ever-04,Leon-05,Tzoum-06a,Tzoum-07,Kaz-07,Harm-09,Fotein-06}, 
CLF potential \cite{Shan-05,Fotein-06}, tube potential \cite{Zhou-06}, 
dilution exponent \cite{Shan-06}, onset of entanglements \cite{Tzoum-09a,Tzoum-09b}, 
and even the literal picture of a tube \cite{Steph-10,Steph-11a,Steph-11b,Baig-10a,Baig-10b} 
enclosing a chain and the associated tube survival probability, have been investigated. 
Despite the available microscopic information, the many chain nature of entanglement remains unclear.
A possible reason is that most of the simulations follow tube model ideas and 
examine, with the help of PPs, single chain properties. Nevertheless, it is the many chain nature of entanglement 
which is elusive and is replaced by a mean field. Therefore, rather than examining single chain properties, which 
are the result of tube confinement, we can try to analyze the {\it chain environment} that 
gives rise to confinement. This is the strategy followed here. 

In order to make the problem tractable, we have only examined systems with trapped entanglements; 
a polyethylene (PE) melt where all chain ends are grafted to other chains, and a perfect PE network 
formed by connecting cross-links initially situated at the sites of a diamond lattice. In this way, our 
results are free of complications related with CR processes. The latter can be addressed after a suitable 
microscopic description of entanglement has been given. In the overview of the paper given below,
the terms tube and tube constraint are used as generic terms 
for entanglement constraints, and are not specifically related to tube model ideas.

As probes of the entanglement environment, we have used the PPs themselves. 
Exploiting the fact that they sample the uncrossability constraints of the surroundings, 
we show that, by mapping short molecular dynamics (MD) trajectories to PP trajectories, 
it is possible to detect all the pairwise chain interactions
assembling the tube constraint. To verify that these interactions result from true topological constraints, 
we ensure that they satisfy a local topological criterion. The latter is based on the common idea of
two concatenated rings that define a {\it binary link}. Therefore, entanglement constraints are decomposed 
into a large set of {\it local links}, which is more or less complete. 
Similar ideas have been explored, analytically, by Iwata and Ewards \cite{Iwa-88,Iwa-89}.

By construction (chain shrinking), all these links are sampled inside the pervaded volume of a chain.
The mechanism by which link sampling takes place is clarified in detail. 
While some of those links are continuously present, many of them are intermittent or `blinking' links.
They are not sampled continuously over a specific observation time.  
Less constraining (weak) links are sampled less frequently. This leads to a distribution of linking times,
and thus the links assembling the tube constraint have a spectrum of confinement strengths, 
with maxima at the strong~\cite{Wang-11} and weak part of the spectrum. 

From this picture, we then develop a simple statistical model and we introduce a pairwise parameter,
that can quantify the pairwise versus collective character of mean-field type links 
representing many uncrossability interactions. With the help of these tools we show quantitatively that; 

$\bullet$~~~~the entanglement environment consists of a plethora 
             of pairwise interactions, so that it is inherently collective; 

$\bullet$~~~there exist strong and weak pairwise uncrossability constraints; 

$\bullet$~~~~the majority of pairwise interactions are weak and confinement is mainly enforced by
             the strongest ones; 

$\bullet$~~~~weak topological interactions can be fully relaxed by stored length fluctuations, 
             and they do not contribute to the low frequency storage modulus of an elastomer, or the 
             plateau modulus of a melt; 

$\bullet$~~~~the average effect of the collective entanglement environment
             is to confine a chain through a much smaller number of mean-field type links,
             which have a dominant pairwise character; 

$\bullet$~~~~the pairwise parameter is chain length and entanglement density independent.

In order to be able to map the entanglement environment to specific pairwise interactions, we present a 
procedure for selecting the strongest links according to a given link density. The procedure introduces 
an auxiliary time scale which is self-consistently determined to be approximately one $\tau_{\rm e}$. 
It is shown that strong links have the property that they are sampled at least once within this time
interval. By following the space-time trajectories of selected links, under equilibrium, deformation, 
or flow, we could obtain useful mesoscopic information for the development of hierarchical slip-link, 
or single particle \cite{Kindt-07,Briels-09,Padd-11} simulation schemes. The mean square displacement 
of selected links is presented in the appendices.

The PPs employed in our study have been generated by the CReTA algorithm \cite{Tzoum-06a}, 
which applies a contour reduction (chain shrinking) scheme. 
Different schemes, based on time averaging \cite{Read-08,Miln-11a}, have been proposed. 
However, the presented results are not based on specific properties of the PPs, except  
that they probe pairwise uncrossability constraints in the pervaded volume of a chain. 
Thus, our conclusions are fairly general and independent of the particular 
scheme employed for obtaining the PPs. 

In order to make the article more readable and focus on the analysis of entanglement
constraints, many of our results are presented in the Appendices.
Our microscopic description is also supported by several {\it explanatory videos \cite{Tzoum-Vids}}
provided with the Supporting Information for this article. 

\section{Systems Studied}
\label{section:Systems}

The systems analyzed are a PE network "rubber" and an end-grafted PE melt (see next section).
The rubber system contains 64 tetrafunctional cross-links and 128 chains. 
The cross-links are initially situated at the sites of a diamond lattice and all chains are
attached to cross-links at both ends. 
Thus, the network is {\it{perfect}} (absence of pendant chains, chain loops, etc.). 
Each chain consists of 201 united atoms, not counting the cross-links at its two ends. 

The melt system consists of 16 linear chains of 500 united atoms each. In order to examine 
the chain length dependence of some of our results, we also analyzed a melt 
with eight chains and double the chain length, i.e., of 1000 united atoms.
The entanglement density of these melt configurations is approximately the same.
Whenever required, these systems will be denoted by (C500) and (C1000),
but in general, our analysis will focus on the (C500) system. 

The simulations were performed in the NPT ensemble, at temperature 450 K, pressure 1 Atm,
by employing the united atom TraPPE\cite{Trappe-98, Trappe-99} force field and 3D
periodic boundary conditions.
United atom beads correspond to $\rm{CH_2}$ methylene units, except chain ends, 
trifunctional, and tetrafunctional cross-links, which correspond to $\rm{CH_3}$, $\rm{CH}$ units, 
and carbon atoms, respectively.

The rubber was equilibrated by the gradual push-off\cite{Auhl-03} method, discussed in Appendix I. 
The melt was equilibrated by standard connectivity altering\cite{Kara-02a,Kara-02b} Monte 
Carlo algorithms. The shortest relaxation time related with the tube constraint, $\tau_{\rm e}$,
is 1.3 ns for the rubber, and 1.7 ns for the melt (C500). 
It was determined from the first crossover time in the mean square displacement (msd) of midchain 
monomers (Appendix II). The longest relaxation time of each system, $\tau_{\rm R}'$, is determined from the msd crossover 
to an invariant regime (absence of long-range diffusion), and is 9.8 ns for the rubber and 11.5 ns 
for the melt (C500). The corresponding crossover times for the (C1000) melt are 1.6 and 
19.4 ns. Details about the dynamics and the equilibration of the systems will appear 
elsewhere\cite{Moro,Stef}.  

Both systems share, approximately, the same density~\cite{Fotein-09,density}, $\rho \simeq 0.78$ g/cm$^{3}$,
and characteristic ratio, $C_{\infty}$. The latter is $8.5$ for the rubber and $8$ for the (C500) melt. 
Volumetric, structural~\cite{Kara-02a,Kara-02b}, 
and entanglement~\cite{Tzoum-06a} properties of the melt system have been presented elsewhere.
They are in very good agreement with corresponding experimental estimates. 
The rubber system is structurally similar to the melt, except for entanglement properties. Its 
$M_{\rm e}$ was estimated~\cite{Tzoum-06a} from the Kuhn segment of the PPs, and is about half 
that of the PE melt. The large entanglement density of the rubber is also
reflected \cite{Moro} in a comparatively high, entanglement-dominated, shear modulus for an elastomer, 
of 5.9 MPa. As is discussed in Appendix I, it is due to the short length of the precursor
chains and the artificial preparation conditions of the network.

\section{Methodology}
\label{section:Method}

For both systems, we generated a molecular dynamics (MD) trajectory, which is of length
$35 \tau_{\rm e}$ for the rubber and approximately $50 \tau_{\rm e}$ for the melts, 
covering a few times the longest relaxation time of each system.
Our purpose is to use the trajectory as an ensemble of states with invariant topology,
representative of the entanglement constraints in the system.
This means that under thermal motion all entanglements should remain 
trapped, a condition satisfied in the perfect network but not 
in the melt, due to the free motion of chain ends.

The melt topology can become invariant by introducing additional topological constraints.
Our choice is to graft (bond) all chain ends to their nearest united atom bead along a different 
nearby chain. Thus, the melt system will be referred to as {\it{end-grafted melt}} (EG-melt).
Effectively, the EG-melt becomes a perfect trifunctional network with 
as many cross-links as chain ends. The initial chains are grafted at both ends
and have on average two randomly placed cross-links along their length
(from the grafted ends of other chains).

In a MD simulation of this network, the conformations sampled by grafted chains are restricted
by the same topological constraints present in the parent melt state, plus the constraints 
due to grafting. From such a network trajectory we store chain conformations by excluding 
all bonds inserted to impose grafting, i.e., the network trajectory is mapped to 
a melt system, with virtually free chain ends. Thus, starting from a melt configuration
with a specific topology, we can generate thermally fluctuating chain conformations with approximately 
the same topology, by eliminating CR through grafting.

Our results for the EG-melt refer to this `perturbed' MD trajectory.
The perturbation preserves the average PP contour length, but it alters
contour length fluctuations. Because of the presence of additional constraints, the grafted chains
cannot sample certain Rouse modes with long length scales. However, the short and medium length scale modes, 
which are responsible for the sampling of the enclosing tube, are adequately sampled.
In this respect, at short times, the dynamics of grafted chains are similar to those
of melt chains with free chain ends. For example, at short times, the mean square displacement 
(see Appendix II) scales as $t^{0.5}$ and then displays a crossover to a 
regime scaling as $t^{0.28}$, a crossover indicative of tube confinement.



\section{Reduction to Networks of Local Links}
\label{section:Reduction}

The generated MD trajectories were subjected 
to topological analysis, frame by frame, by using the CReTA~\cite{Tzoum-06a} 
(Contour Reduction Topological Analysis) algorithm.
CReTA reduces a dense system of polymer chains to the corresponding 
system of primitive paths (PPs), constructed as the shortest paths~\cite{Doi-86}
under the same `topological constraints' (TCs) as the original
chains. By fixing chain ends in space and by prohibiting
chain crossing, the contour lengths of all chains are simultaneously 
minimized (shrunk), until they become piecewise linear objects
coming together at the nodal points of a network. 
During contour reduction, chain thickness is progressively reduced. 
Finally, the PPs become very thin objects consisting of {\it{fused beads}}
which can be also mapped to the initial chain monomers.
Explanatory videos and other details can be found elsewhere~\cite{Tzoum-06a, Tzoum-Vids}.

Network nodes are placed where PPs are mutually blocked (entangled)
and contour reduction can no longer proceed (see network pictures in Supporting Information).
They represent effective spatial localization points of the TCs, that each chain is 
subjected to. They can also be envisaged as local links that have a physical resemblance
to binary slip links. The details of mapping a system of `tightened' entangled curves,
consisting of consecutive beads, as our system of PPs here, 
to a network with point-like nodes and rectilinear edges, will be described 
elsewhere~\cite{Stef}. It will be shown that the mapping is
very stable with respect to changes in the free parameters involved.
A short account of the mapping is given in~\ref{fig:links}, and in Appendix III. 

\begin{figure} [!tbp]
\begin{center}
  \includegraphics[clip,width=0.5\linewidth] {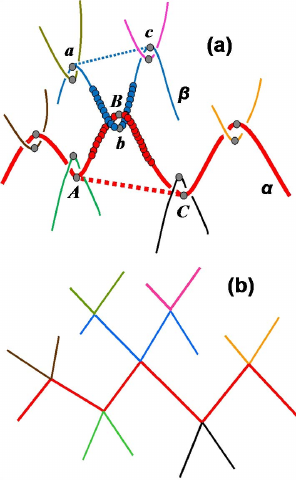}
\end{center}
\caption{\label{fig:links}  Network nodes, local links, and the {\it{topological 
criterion}} satisfied by local links. (a) PPs are composed of consecutive beads. 
Parts $ABC$ and $abc$, of chains $\alpha, \beta$, respectively, constrain each other
and are sketched with a bead structure. All other parts are sketched with contour
lines. The beads which `carry' a TC are colored gray. They mark the points along the
PPs where a suitably defined curvature~\cite{Stef} shows a local maximum. 
Each TC bead is pairwise associated with a TC bead along the PP of a mutually constrained chain. 
Thus, chain ends and TC bead pairs partition a chain into consecutive strands. 
In order to decide if a TC bead pair defines a local link we apply a 
topological criterion at the {\it{bead level}}. For each pair, such as $(B,b)$ between
chains $\alpha,\beta$, we construct the composite strands $ABC$ and $abc$. The ends of these strands 
are virtually connected, and then we check if the strand $ABC$ crosses the area 
enclosed by the virtual segment $ac$ and the beads along $abc$. Similarly, we examine if 
the strand $abc$ crosses the area enclosed by the virtual segment $AC$ and the beads along $ABC$. 
In this way, we examine twice if the $ABC$ and $abc$ {\it{rings}} concatenate.
If at least one of these checks is successful, the TC bead pair $(B,b)$ is promoted
to a network node {\it(local link)}, otherwise it is discarded. Nodal coordinates are calculated from 
the vector average of the coordinates of each TC pair. Hence, the configuration shown in (a) becomes the
one shown in (b), and the underlying topology is mapped to a network. Dynamical videos of such networks, and 
realistic closeups on PP networks of a PE melt can be found in the Supporting Information.}
\end{figure}

\section{Results and Discussion}

\subsection{Fluctuations in the Number of Local Links and Linked Chains (LCs)}
\label{section:Links}

\begin{figure} [!tbp]
\begin{center}
  \includegraphics[clip,width=0.8\linewidth] {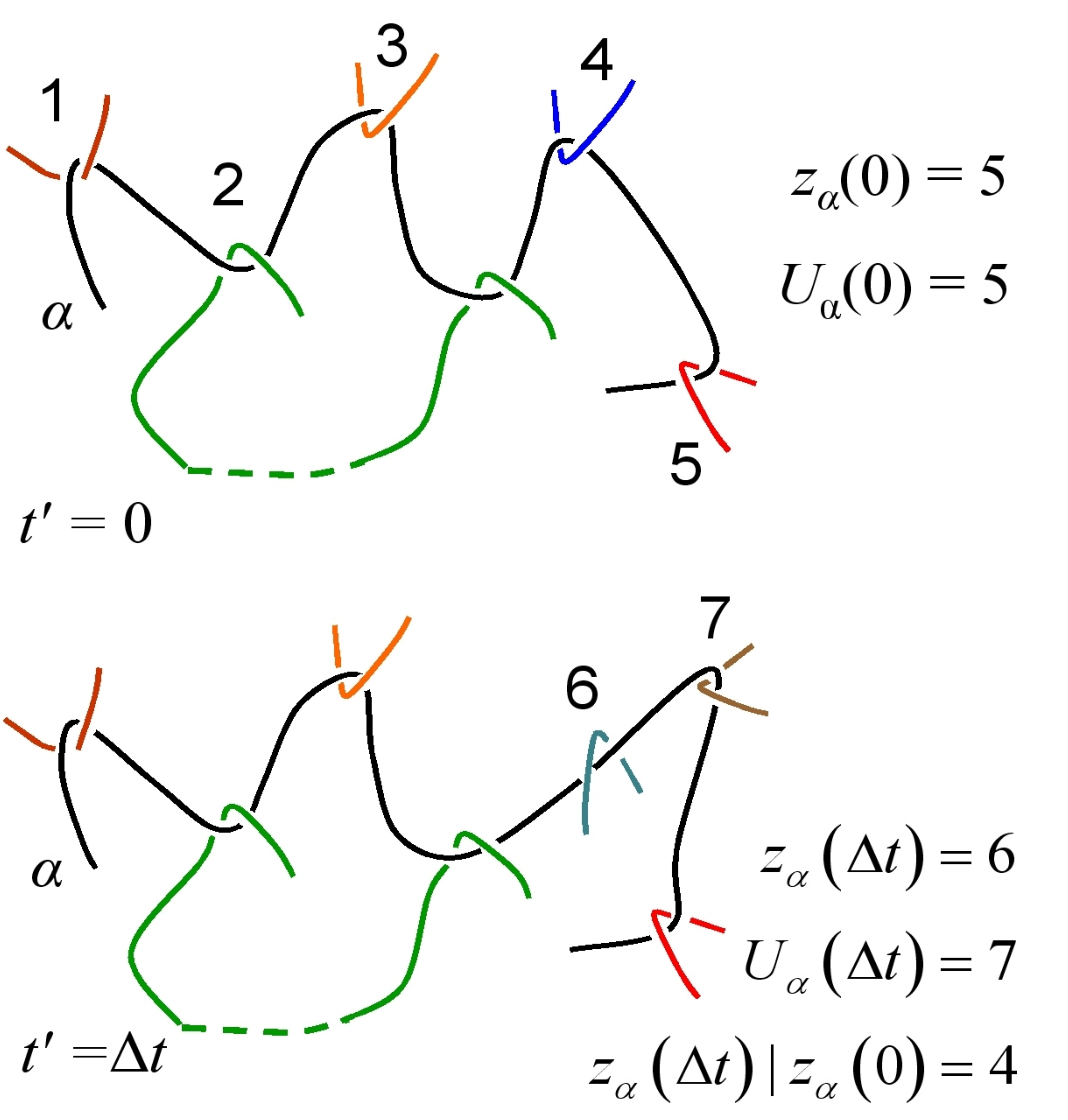}
\end{center}
\caption{\label{fig:zuq} 
 At times zero and ${\Delta}t$, chain $\alpha$ is linked to a group of five and six different chains, respectively.
 Since four of these chains are the same, the overlap between these groups is $z_{\alpha}({\Delta}t)|z_{\alpha}(0)=4$.
 The cumulative number of different chains sampled by $\alpha$ within ${\Delta}t$ is 
 $U_{\alpha}({\Delta}t)=7$. By definition $U_{\alpha}(0) = z_{\alpha}(0)$. For the chain lengths studied,
 the probability that two chains are entangled with more than one link, as chain $\alpha$ and chain 2 here,
 is less than 10\% (see Appendix IV). Thus, throughout the document we will consider that, at any time $t$,
 the number of links of chain $\alpha$ is equal to the number of linked chains $z_{\alpha}(t)$. }
\end{figure}

Here, we examine the tube constraint at the level of the surrounding 
chains that create it. To this end, we define as $z_{\alpha}(t)$ the number of 
chains linked, {\it{instantaneously}}, to a specific chain $\alpha$,
at a specific time $t$. As shown in~\ref{fig:zuq}, {\it{all chains in this set are different}}.
That is, if at time $t$ chain $\alpha$ is linked twice or more with chain $\beta$, 
at different places along their contours, then $\beta$ counts as one chain in $z_{\alpha}(t)$,
and vice versa. We will refer to this time-dependent set as the mate chains, or Linked Chains (LCs) 
of $\alpha$. If, due to periodic boundary conditions, $\alpha$ is also linked with different `images' of 
$\beta$ or of itself, then these images 
    \bibnote{Extreme care should be taken when dealing with different chain images. 
    For readers not familiar with simulations, we note that the chain parts which are
    inside the simulation box can belong to different chain images. Here, 76 and 329 
    different chains pass through the simulation box of the EG-melt (C500) and rubber system, 
    respectively. The results are not influenced by periodic boundary conditions.}
are considered as different chains in $z_{\alpha}(t)$. 
Entanglements or knots of $\alpha$ with itself (same image) are not considered. For the 
chain lengths studied here their occurrence is negligible \cite{Suku-05,Tzoum-06a}.
We also note that $z_{\alpha}(t)$ characterizes a chain's entanglement
environment by keeping track of the identities of linked chains at time $t$, in a form
suitable for comparisons at different times. 


In Appendix IV (\ref{fig:p_n}), we show that, when two chains are linked, the
probability that they are coupled with more than one link is very small (less than 10\%).
Hence, the simplification of dealing with linked chains instead of 
individual links is justified. This choice has been made in order to facilitate the 
presentation and the interpretation of our results. We could as well work with individual links.
Therefore, in the following sections {\it{we will use interchangeably the terms
local links and linked chains}} for $z_{\alpha}(t)$, though in our
calculations $\langle z_{\alpha}(t) \rangle$ refers to LCs and not individual links. 

\begin{figure} [!tbp]
\begin{center}
  \includegraphics[clip,width=0.7\linewidth] {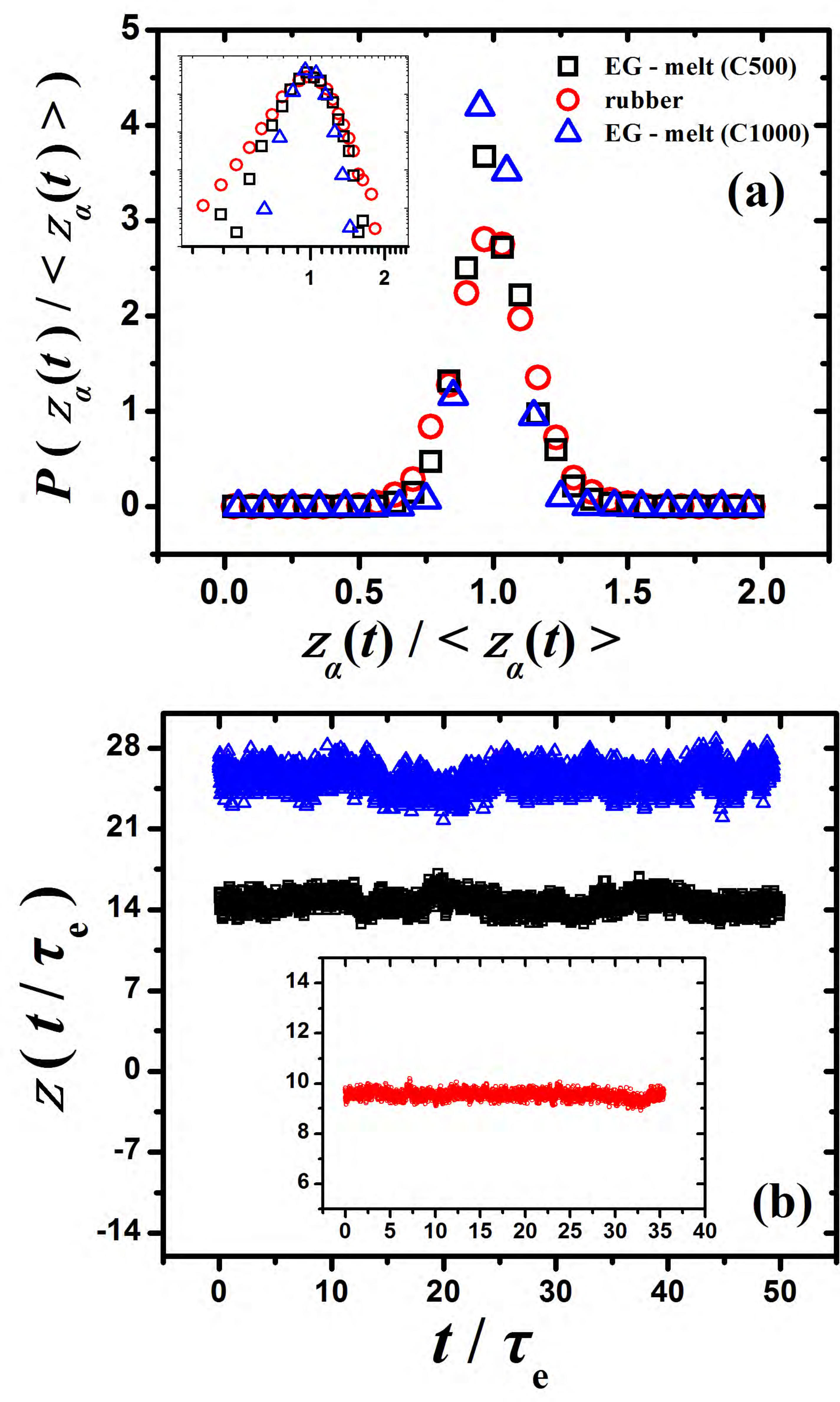}
\end{center}
\caption{\label{fig:z_t} 
(a) Distribution of $z_{\alpha}(t)$  reduced by the corresponding time average 
    $\langle z_{\alpha}(t) \rangle$ (see text). Log-log coordinates in the inset.
    To obtain these results, a separate distribution is constructed for each chain first 
    and then an average is taken over the distributions of all chains. This is done because the entanglement 
    environment (number of entanglements) and related fluctuations may be different for each chain. 
    Thus, we consider each chain as a subsystem, we analyze the fluctuations within each subsystem, 
    and then we average over the subsystems. 
    This is necessary in order to map the fluctuations to a single chain model.
(b) Average number of links per chain at time $t$.}
\end{figure}

In~\ref{fig:z_t}a we plot the distribution of $z_{\alpha}(t)$ reduced by its equilibrium 
time average $\langle z_{\alpha}(t) \rangle$. In contrast to what one would expect for 
a system with solely trapped entanglements, $z_{\alpha}(t)$ fluctuates around a time average value.
In a corresponding slip-link system, $z_{\alpha}(t)$ would be constant. This happens because 
each PP is a {\it dynamically fluctuating object} that samples its environment
by colliding with the contours of the surrounding PPs penetrating its pervaded volume.

Such `collisions' lead to the formation of local links which are not permanent. Instead they
are formed for certain times, i.e., they exhibit various {\it linking times}. However, there 
also exist links which restrict continuously two fluctuating PPs and exhibit a permanent character.
As a result, the instantaneous number of links sampled by a PP, $z_{\alpha}(t)$, fluctuates around
an average value. In Appendices V and VI, and in the videos provided with
the Supporting Information, this issue is discussed in more detail. 
In the inset to~\ref{fig:z_t}a, we also see that the amplitude of these fluctuations
seems to be smaller for long chains. 

The chain average of $z_{\alpha}(t)$ is defined as

\begin{align}
    z(t)  =\frac{1}{N_{\rm ch}}\,\,\sum\limits_{\alpha =1}^{N_{\rm ch}} z_{\alpha}(t)~~, \label{eq:D}
\end{align}

where $N_{\rm ch}$ is the number of chains. In~\ref{fig:z_t}b it is plotted as a function
of reduced time, $t/\tau_{\rm e}$. We see that for both systems it is approximately constant.  
Its time average, ${\langle{z(t)}\rangle}$, is $9.5$ for the rubber, and $14.6$, $25.3$
for the EG-melt (C500) and (C1000), respectively.  
These values lead
    \bibnote{By estimating the ratio of chain length over $(\langle z(t) \rangle + 1)$, which is 
             proportional to the entanglement molar mass, we can compare the link density 
             between the two systems. In the rubber it is about twice that of the EC-melt. 
             Estimating $M_{\rm e}$ from the Kuhn segment of the PPs we are led to 
             the same conclusion. Note that the link density of our systems is larger than the rheological
             entanglement density, as discussed in~\ref{section:Discussion}.}
to a rubber entanglement density which is approximately twice that of the EG-melt.

The fluctuations around ${\langle{z(t)}\rangle}$ are much smaller for the rubber because 
chain averaging is more effective in smoothing individual chain fluctuations.
The rubber consists of eight (sixteen) times more chains than the EG-melt (C500) and (C1000) systems, 
respectively. In a very large system, $z(t)$ would appear as a thin horizontal line, i.e., it would remain 
constant over time. Thus, despite the fluctuations in the number of links of individual chains,
when viewing the system as a single chain (through ${\langle{z(t)}\rangle}$), we observe the behavior 
expected of a slip-link system.  

%

\subsection{Cumulative Number of Sampled links: A Complete Set}
\label{section:Cumul}

Because of the absence of CR, after a certain time the PP cannot sample any new links in its surroundings. 
This fact is exploited in order to gather all sampled links into a {\it complete set} of pairwise 
uncrossability interactions characterizing the entanglement environment.

To elucidate these ideas further, we introduce $U_{\alpha}(t)$, the cumulative 
number of different chains which have been linked to a given chain, $\alpha$, 
up to time $t$. All chains in $U_{\alpha}(t)$  are {\it different}. 
They were not necessarily linked to $\alpha$ at all times between $0$ and $t$. Some of them
may have been linked for a long (discontinuous or continuous) subinterval, 
some of them for just one instant, or for the whole time $t$. As shown in~\ref{fig:zuq},
$U_{\alpha}(t+{\Delta}t)$ is larger than $U_{\alpha}(t)$ by the number of mate chains
which link to $\alpha$, {\it for the first time}, at any instant between 
$t$ and $t+{\Delta}t$. 

\begin{figure} [!tbp]
\begin{center}
  \includegraphics[clip,width=0.8\linewidth] {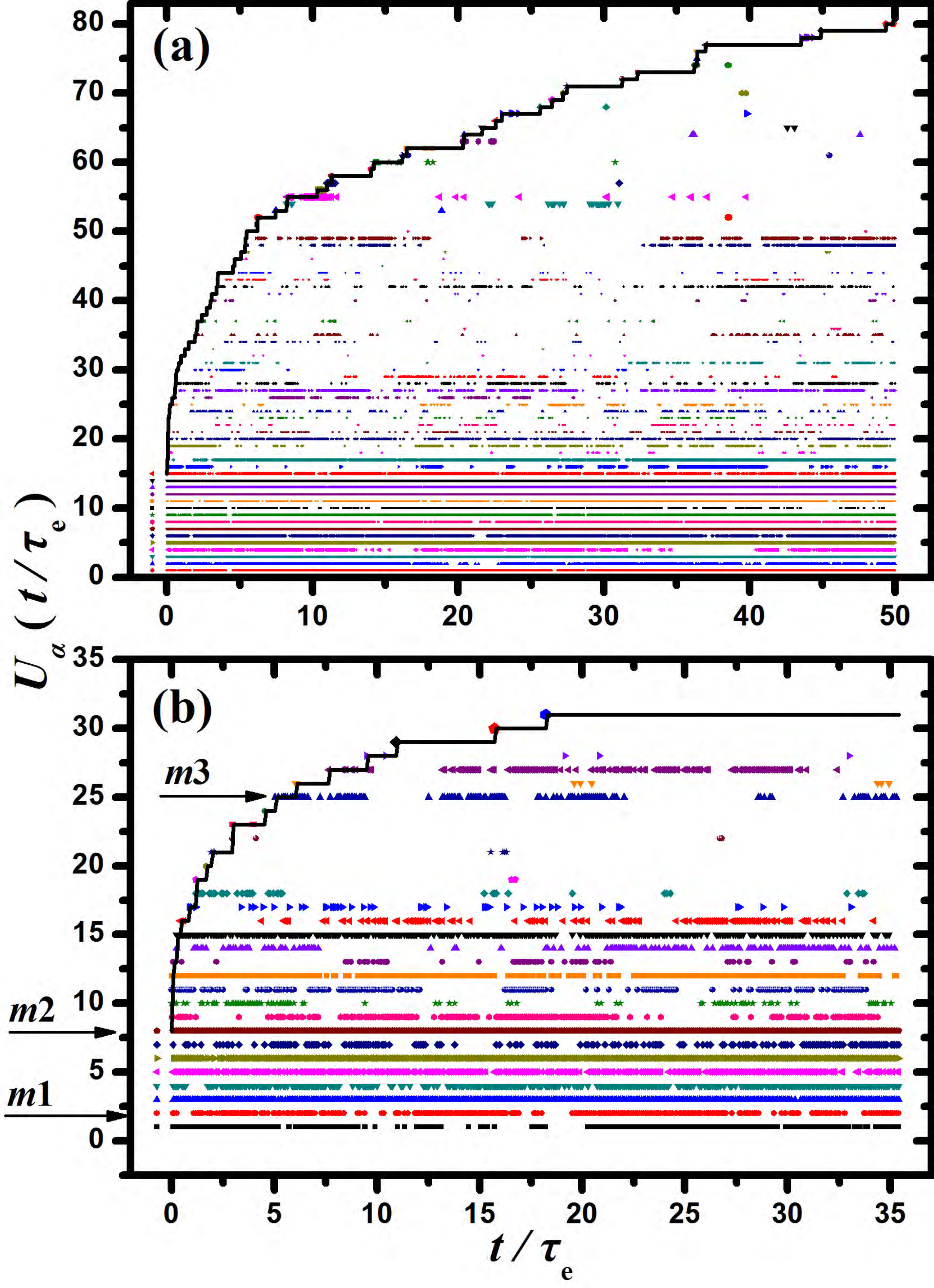}
\end{center}
\caption{\label{fig:U_a}
Cumulative number of different links or linked chains (LCs), $U_{\alpha}(t)$, sampled by a given chain 
$\alpha$ within observation time, $T$, as a function of reduced time, $t / \tau_{\rm e}$. 
(a) EG-melt (C500), $T \simeq 50\tau_{\rm e}$. 
(b) Rubber, $T \simeq 35 \tau_{\rm e}$. 
$U_{\alpha}(t)$ is a stepwise, increasing, integer valued function, shown by a solid line. 
Here, it enumerates with unique integer indices, $i$, cumulatively, the links sampled by chain 
$\alpha$, up to time $t$. When a link, $i$, is found to be paired to $\alpha$ at time $t$, a point with  
coordinates $(i,t)$ is plotted. Therefore, horizontal lines, continuous or segmented, correspond to 
individual `link trajectories', as those denoted by $m1$, $m2$ and $m3$. 
The links sampled {\it for the first time} at a time $t'>0$ lack 
point entries for earlier times $t<t'$ (link $m3$).
For convenience, the links sampled at $t=0$ are also plotted as a separate column to the left of $t=0$. 
By construction, the number of points in a $t$-column is equal to $z_{\alpha}(t)$, and 
$U_{\alpha}(0) = z_{\alpha}(0)$. 
The linking time (LT), $\tau$, of a link, is the total plotted length of the corresponding trajectory.
The reduced linking time, $\tilde{\tau} = \tau / T$, is a $T$-independent quantity characterizing
the {\it `persistence'} of each link, with values $0 < \tilde{\tau} \le 1$. 
In the videos provided as Supporting Information, intermittent link trajectories 
with $\tilde{\tau} < 1$, e.g., link $m1$, appear as {\it `blinking' links} 
and they have various, intermediate confinement strengths. Link trajectories 
with $\tilde{ \tau} \simeq 1$, e.g., link $m2$, correspond to permanent, {\it strong links}.
There also exist links with $\tilde{\tau} \simeq 0$, e.g., link $m3$, which are very {\it weak links}. 
Similar figures for other chains can be found in the Supporting Information.
}
\end{figure}

An informative graph of $U_{\alpha}(t)$, together with individual `link trajectories', is shown 
in~\ref{fig:U_a}. In this figure, the systems are observed at equidistant time instants, spaced by 0.01 ns,
over a total observation time (OT), $T$, which covers a few times the longest relaxation time of each system.
Specifically, $T \simeq 35 \tau_e$ for the rubber, $T \simeq 50 \tau_e$ for the EG-melts. 

We see that some links `blink'. They are not present during
the entire OT. From the number of points in a trajectory we can estimate the corresponding linking time 
(LT), $\tau$, i.e., the total time during which a particular link was active. In order to analyze the plethora of 
links we introduce the {\it link persistence}, $\tilde{\tau} = \tau / T$, which is basically the reduced linking
time of a link. As the observation time increases, $\tau$ becomes proportional to $T$, with a 
different prefactor for each link (see Supporting Information). Thus, for adequately long OTs, as the ones utilized here, 
the $\tilde{\tau}$'s are equilibrium $T$-independent quantities that characterize the `persistence' of link trajectories 
with values~~$0 < \tilde{\tau} \le 1$. 


A subset of the links appears to be permanent, with $\tilde{\tau} \simeq 1$.
Most of the LCs sampled at $t=0$ have this property.
Instantaneous sampling at any $t$ (a $t$-column in~\ref{fig:U_a}),
will have a large overlap with this subset, which are basically {\it strong links}.
This is evident in individual $t$-columns, where most of the points are 
located on horizontal lines originating at $t=0$. 
Note that, by construction, the number of points in a $t$-column is $z_{\alpha}(t)$.

A striking feature of~\ref{fig:U_a} is that the majority of the $U_{\alpha}(T)$ links have small, 
even vanishing $\tilde{\tau}$'s. Moreover, the number of links sampled over the whole OT, 
$U_{\alpha}(T)$, is much larger than those sampled instantaneously, 
$z_{\alpha}(t)$, at any time. Furthermore, after a certain time, 
$t'$, most of the links with appreciable $\tilde{\tau}$ have been sampled. 
The links sampled for the first time at later times, $t>t'$, appear to have a $\tilde{\tau} \simeq 0$.

\begin{figure} [!tbp]
\begin{center}
  \includegraphics[clip,width=0.8\linewidth] {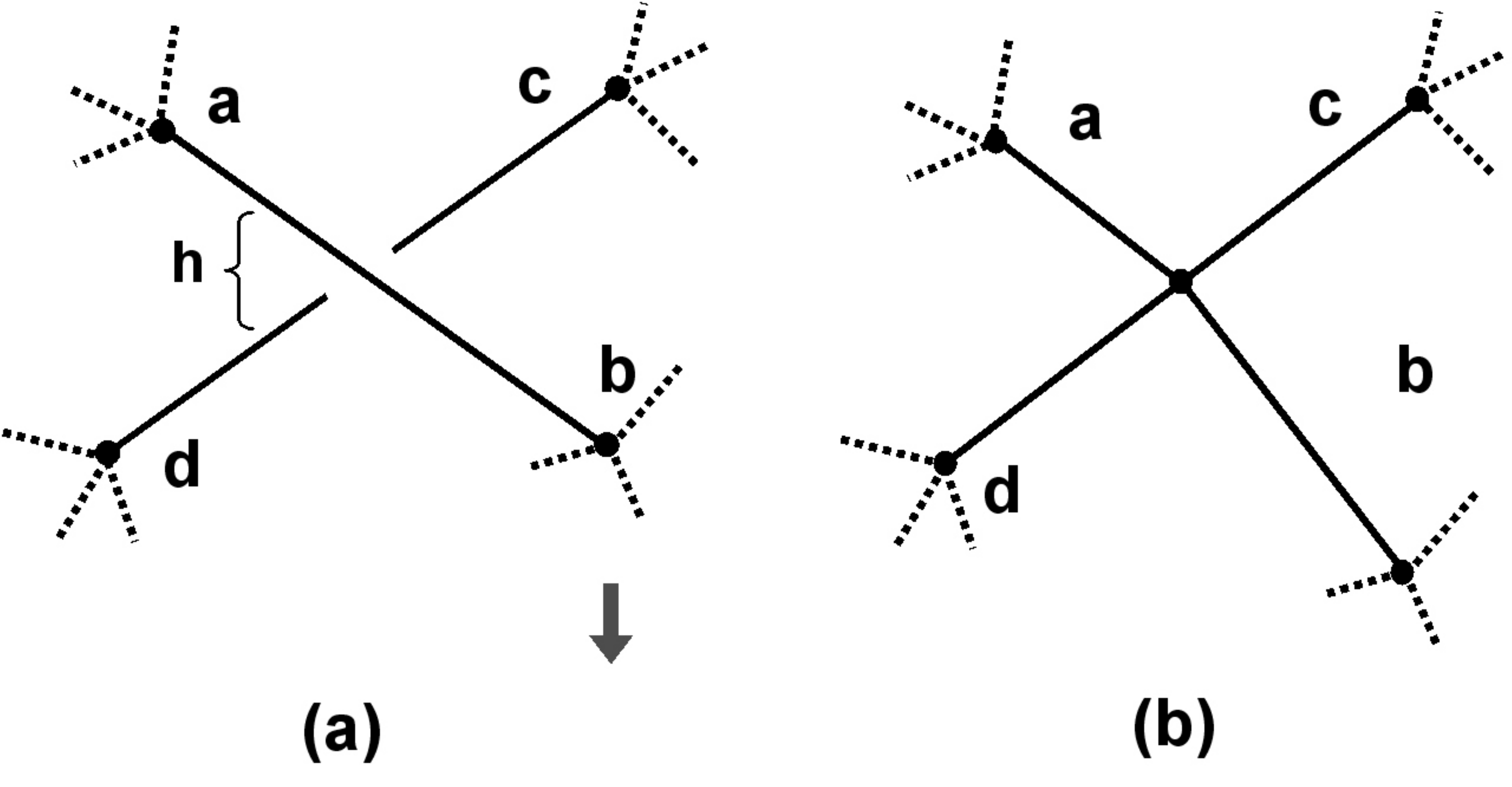}
\end{center}
\caption{\label{fig:hooking} 
Lateral link sampling (LLS) mechanism by which PP sampling of new links
takes place in a spatially fixed tube. 
(a) The instantaneously sampled PP segments are one above the other, at a small distance $h$.
(b) A small downward displacement of one node leads to a local link between the two PPs.
The reverse process is also possible. In reality (see Appendix V), all nodes fluctuate due to the thermal 
motion of the real chains and the associated redistribution of stored length. This leads to lateral PP 
fluctuations and to this kind of sampling. Therefore, such events are frequent. 
Note that the strands in the figure are topologically constrained, whether they form a link or not. When 
$h$ is of the order of lateral PP fluctuations they can become linked for certain time intervals. 
If the {\it persistence}, $\tilde{\tau}$, of such an occasionally formed link 
tends to zero, then the corresponding local pairwise interaction can be considered as weak.}
\end{figure}

The sampling mechanism that generates blinking links is relatively simple, 
and is demonstrated in~\ref{fig:hooking}.
The driving force for this mechanism are {\it stored length fluctuations}, as described in detail in Appendix V.
In order to avoid confusion with constraint release (CR) mechanisms driven by chain end motion, 
we call this mechanism lateral link sampling (LLS).
However, upon nonlinear deformation or flow, LLS could possibly act as a CR mechanism.
A similar mechanism was implemented~\cite{Padd-01a,Padd-02,Padd-11} 
by Padding and Briels to enforce chain uncrossability in a coarse grained model of polyethylene.

%

\begin{figure} [!tbp]
\begin{center}
  \includegraphics[clip,width=0.8\linewidth] {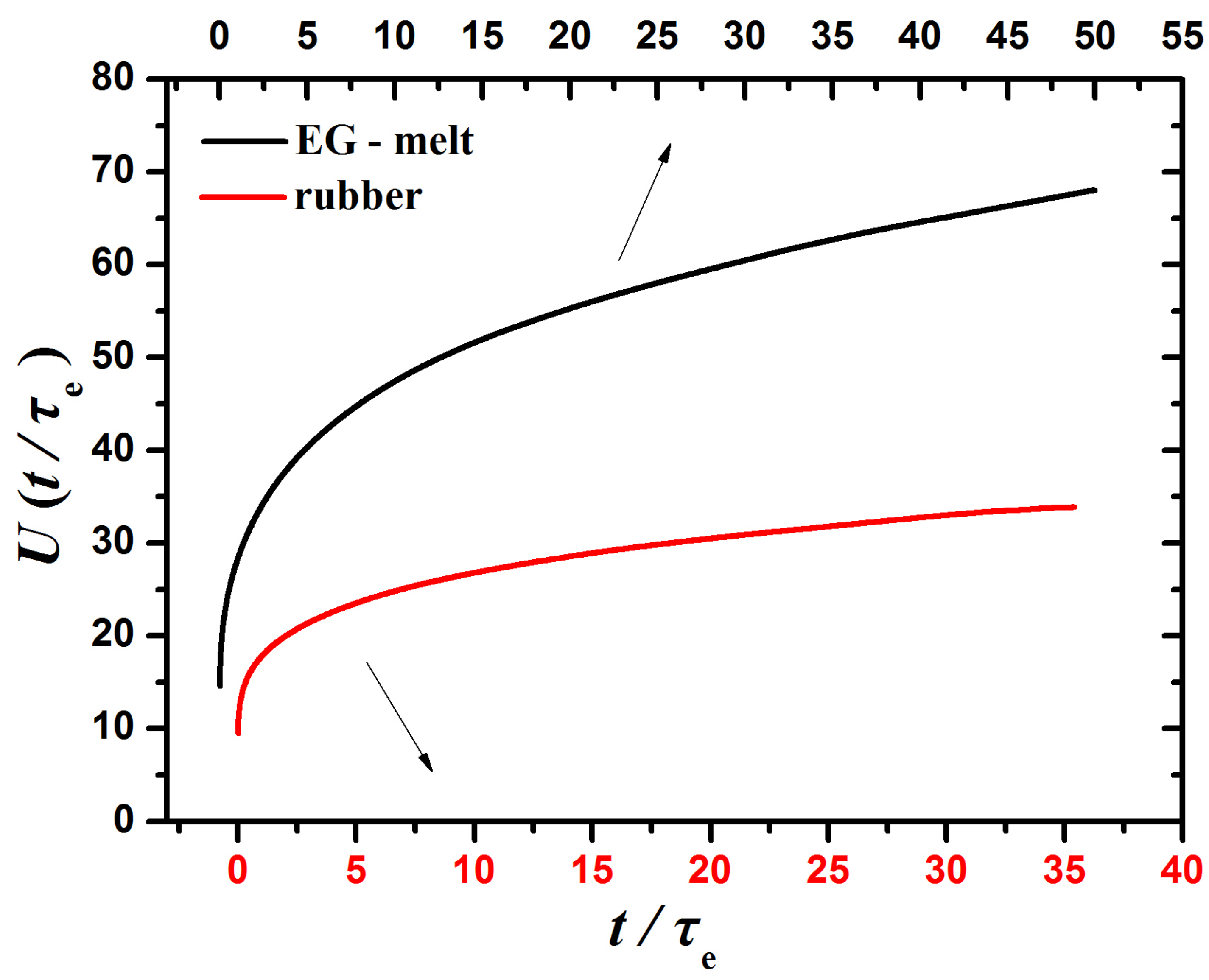}
\end{center}
\caption{\label{fig:U} 
  Average number of $different$ linked chains or links which make the tube constraint up to time $t$.
  $\tau_{\rm e}$ is 1.3 ns for the rubber and 1.7 ns for the EG-melt (C500).}
\end{figure}

The chain average of $U_{\alpha}(t)$ is defined as

\begin{align}
    U(t) =\frac{1}{N_{\rm ch}}\,\,\sum\limits_{\alpha =1}^{N_{\rm ch}} U_{a}(t) ~~, \label{eq:U}
\end{align}

and is shown in \ref{fig:U}. It is estimated by averaging over multiple time origins, so that by 
construction $U(0) = {\langle{z(t)}\rangle}$ (since $U_{\alpha}(0) = z_{\alpha}(0)$). 
We see that $U(t)$ increases rapidly and approaches a plateau which is much larger than the average number
of links ${\langle{z(t)}\rangle}$ per chain (see~\ref{fig:z_t}b). This is more evident in the 
EG-melt (C500); the longer chains and the wider tube of this system lead to a larger number of 
surrounding chains assembling the tube. In this system, $U(t)$ does not seem to level off. It increases 
slowly, with a very small but nonzero decreasing rate. A true plateau with a slightly larger value 
is expected to exist (fixed tubes), at some longer $t$.

However, it is not required to sample this regime, since the links
sampled for the first time at such long times have a vanishing $\tilde{\tau}$ (see~\ref{fig:U_a}). 
Most probably, their sampling involves large, cooperative stored length fluctuations (Appendix V), 
which are not frequent
  \bibnote{On this point, strong evidence is provided by $q(t)$, introduced in Section 5.5, 
  which reaches a plateau at times where $U(t)$ is still increasing.}. 
While these links are part of the entanglement environment of a chain, 
their confinement strength is negligible, and they can be safely considered as {\it weak}. 

Therefore, for the systems studied, and for all practical purposes, we consider that an observation 
time of $T \simeq 40-50 \tau_{\rm e}$ is adequate for sampling all pairwise interactions playing 
a significant role in confinement. We also consider that the links contained in $U_{\alpha}(T)$ form a
{\it complete set}. This set applies to $\alpha$ the reputed tube constraint. 
The state of entanglement of $\alpha$, described by $z_{\alpha}(t)$,
{\it can be projected}, at any time, to the $U_{\alpha}(T)$ links. Corresponding statements can also be 
made for $U(T)$, which views the system as a single chain. In the next sections, we will refer to 
these complete sets with the symbols $U_{\alpha}$, $U$, defined as 

\begin{align}
    U_\alpha = U_{\alpha}(T) ~~,~~ U = U(T). \label{eq:U_a_U}
\end{align}

These are adequate representations of the full sets $U_{\alpha}(\infty)$, $U(\infty)$,
which may have contributions from additional weak links that can be neglected.


\subsection{Number and Weight Distribution of Link Persistences}
\label{section:Distr}

\begin{figure} [!tbp]
\begin{center}
  \includegraphics[clip,width=0.8\linewidth] {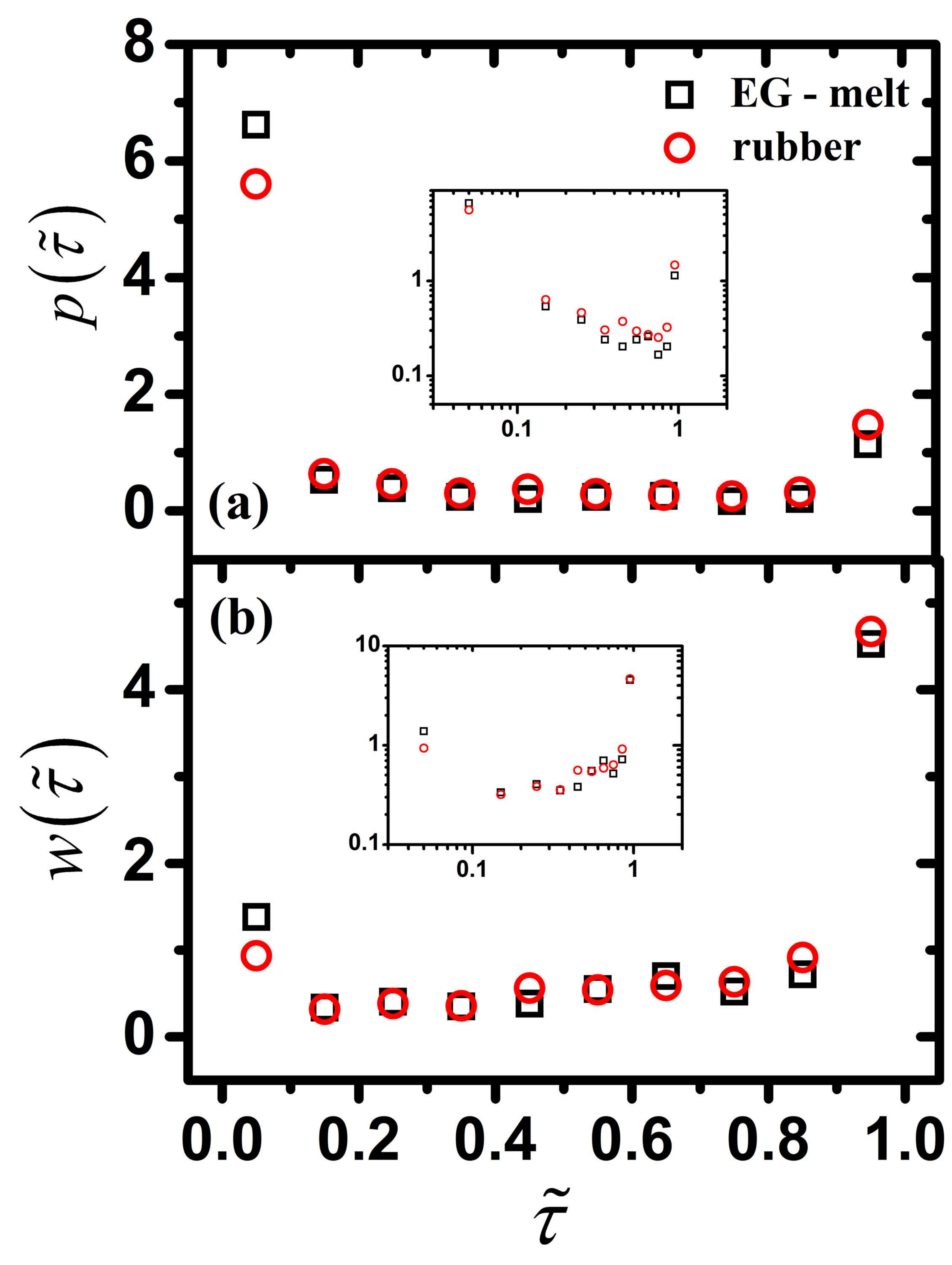}
\end{center}
\caption{\label{fig:pw} (a) Normalized distribution of link persistences. Since all links are equally weighted, 
this distribution corresponds to the {\it number distribution} of $\tilde{\tau}$'s. 
(b) Normalized {\it weight distribution} of link peristences. 
Log-log coordinates in the insets. The observation time, $T$, is $35 \tau_{\rm e}$ for the rubber, 
and $50 \tau_{\rm e}$ for the EG-melt (C500).}
\end{figure}

We have seen that for the average single chain, the entanglement environment consists of a complete
set of $U$ {\it different} links. Here, we characterize this set through the 
distribution of the persistence of links. 
In \ref{fig:pw}a we present the normalized probability distribution, 
$p(\tilde{\tau})$, discretized in intervals of $\Delta\tilde{\tau} = 0.1$. The fraction
of the $U$ links falling in each interval is $p(\tilde{\tau}) \Delta\tilde{\tau}$.
We observe that $p(\tilde{\tau})$ has an absolute maximum at $\tilde{\tau} \simeq 0$, and a second maximum at
$\tilde{\tau} \simeq 1$. Hence, the majority of the $U$ links are weak, and the average link persistence, 
${\tilde{\tau}}_{\rm n}$, is quite small, namely, $0.30$ for the rubber, and $0.24$ for the EG-melt. 
  
In the Supporting Information we present $p(\tilde{\tau})$ for several increasing $T$, 
starting from $T=\tau_{\rm e}$. It is shown that for the observation
times discussed here the distribution becomes invariant. That is, had we increased $T$, we 
would have obtained, practically, the same distribution. This is a necessary condition in order 
to consider $U=U(T)$ as a complete set and $T$-independent. It is satisfied when sampling the
entanglement environment for long OTs, where the linking time of any link becomes proportional to 
$T$, so that the corresponding persistence, $\tilde{\tau} = \tau/T$,
acquires an equilibrium value. Additionally, at long $T$'s the weak links of the environment have been
adequately sampled. 


In order to associate the persistence of a link with a confinement strength, we have to
note that in $p(\tilde{\tau})$ all links are equally weighted. Therefore, ${\tilde{\tau}}_{\rm n}$
corresponds to the number-average link persistence. 
However, equal weighting hides the fact that the tube constraint is applied mainly by links with 
$\tilde{\tau} \simeq 1$. In order to give greater importance to these links we need to estimate the weight
average distribution of $\tilde{\tau}$'s, $w(\tilde{\tau})$, which is defined as 
\begin{align}\label{eq:w}
 w(\tilde{\tau})  = \frac{\tilde{\tau}} {{\tilde{\tau}}_{\rm n}} p(\tilde{\tau})~~~.
\end{align}
In this way, the link of a chain is weighted according to the ratio of its persistence, over the sum of the 
persistencies of all links to this chain. 
For a link between chains  $\alpha$, $\beta$, with persistence ${\tilde{\tau}}_{\alpha \beta}$,  
the corresponding weight is
\begin{equation}
  w_{\alpha\beta} \left( {\tilde{\tau }}_{\alpha\beta} \right) = \frac{{\tilde{\tau }}_{\alpha\beta}} {\sum\limits_{\beta=1}^{U_{\alpha}} {\tilde{\tau }}_{\alpha\beta} } = \frac{ {\tilde{\tau }}_{\alpha\beta} }  {\left\langle z_{\alpha}(t) \right\rangle} \label{eq:weight}
\end{equation}
and corresponds to the {\it confinement strength} of this link.
The closer $\tilde{\tau}$ is to unity, the stronger the confinement provided by a link.
Similarly, $w(\tilde{\tau}) d\tilde{\tau}$ is the fraction of confinement provided by the $U$ links having
a persistence between $\tilde{\tau}$ and $\tilde{\tau} + d\tilde{\tau}$. 
Note that $p(\tilde{\tau})$ and $w(\tilde{\tau})$ view the system as a single chain.

The normalized weight distribution $w(\tilde{\tau})$ is shown in~\ref{fig:pw}b.
We see that confinement comes from, mainly, links with $\tilde{\tau} \simeq 1$. 
However, the form of $w(\tilde{\tau})$ does not allow for a separation of links into
strong and weak. This issue is discussed in detail in \ref{section:Plateau}.
The {\it weight-average} link persistence, ${\tilde{\tau}}_{\rm w}$, is $0.76$, for both systems.
It is much larger than ${\tilde{\tau}}_{\rm n}$ and it corresponds to the  
strength weighted average persistence of links.

In order to clarify this section further, it is useful to make an analogy
with a corresponding slip-link system where all entanglements are trapped.
In this case the links do not blink and all link trajectories are continuous. 
Then, assuming $Z$ links per chain, we have $z(t) = Z$, at any $t$, and therefore
$U = \langle z(t) \rangle = Z$, for any observation time.
For each link ${\tilde{\tau}} = 1$ (permanent links), which leads to ${\tilde{\tau}}_{\rm n} = 1$. 
Regarding $p(\tilde{\tau}) d\tilde{\tau}$, 
it would be zero for $\tilde{\tau} \neq 1$, and unity for $\tilde{\tau} = 1$ 
(delta function at $\tilde{\tau} = 1)$. 
Regarding $w(\tilde{\tau})$, from~\ref{eq:w} we have $w(\tilde{\tau}) = p(\tilde{\tau})$,
and thus, ${\tilde{\tau}}_{\rm w} = {\tilde{\tau}}_{\rm n} = 1$. 
Concerning individual links, from~\ref{eq:weight} we find that all links have an equal 
confinement strength, $w = 1/Z$.

Finally, we note that the strength of a link could be associated with additional quantities.
For example, a time average of suitably defined elastic interactions over the four strands making the link,
the local friction between the mated chains, the real space fluctuations of the link, 
the monomer diffusion through a link, etc.
At a first approximation, here, we consider that all these quantities are expressed indirectly 
through the link `persistence'. They could be extracted from atomistic simulations
as long as individual link trajectories can be defined, as discussed in \ref{section:Selection}
and in Appendix II). 

\subsection{Collective Entanglement Environment, and Mean-Field Type Links}
\label{section:BinCol}

A useful relation can be obtained by examining~\ref{fig:U_a}, again.
Denoting by $\tau_{\alpha\beta}$ the LT of a link between chains 
$\alpha$, $\beta$, the following equation holds, 
\begin {equation}
  \sum\limits_{\beta =1}^{{{U}_{\alpha}}}{{{\tau }_{\alpha \beta }}}=\int\limits_{0}^{T}{{{z}_{a}}}(t)dt~~, \label{eq:sumA}
\end {equation}

where $U_\alpha = U_\alpha(T)$. With reference to~\ref{fig:U_a},
\ref{eq:sumA} states that in order to add up the LTs of the links sampled by $\alpha$,
we can sum either segmented horizontal lines (left-hand side), or the number of points 
in vertical $t$-columns (right-hand side). In either case, the sums correspond to the total 
plotted length of the link trajectories, up to time $T$.


To proceed further we note that the average LT of a link of chain $\alpha$, 
${{\tau}}_{{\rm n},\alpha}$, and ${\langle z_{\alpha}(t) \rangle}$, 
can be written as 

\begin{equation}
   {\tau_{\rm{n},\alpha}} = \frac{1}{U_\alpha} \sum\limits_{\beta =1}^{U_\alpha} {\tau_{\alpha \beta}} = T {\tilde{\tau}}_{{\rm n}, \alpha}~~,  \label{eq:tau_DD}
\end{equation}

\begin{equation}
   \left\langle {{z}_{\alpha }}\left( t \right) \right\rangle = \frac{1}{T}\int\limits_{0}^{T}{{{z}_{a}}}(t)dt~~. \label{eq:tau_D}
\end{equation}


Then, by dividing~\ref{eq:sumA} with $U_{\alpha}T$, we find

\begin{equation}
   U_{\alpha} {{\tau}}_{{\rm n},\alpha} = {\langle z_{\alpha}(t) \rangle} T~~.  \label{eq:rule_a1}
\end{equation}

Since $U_{\alpha}$ and ${{\tau}}_{{\rm n},\alpha}$ are independently distributed
variables
   \bibnote{A chain average over these quantities satisfies the relation
            ${\langle U \tau_{\rm n} \rangle} = {\langle U \rangle} {\langle \tau_{\rm n} \rangle}$.}
over the chains of the system (numerically checked), the chain average
of \ref{eq:rule_a1} reads 

\begin{equation}
   U {{\tau}}_{\rm n} = {\langle z(t) \rangle} T ~~,  \label{eq:rule_a}
\end{equation}

that can also be written in the $T$-independent form 

\begin{equation}
   U {\tilde{\tau}}_{\rm n} = {\langle z(t) \rangle} \cdot 1 ~~,  \label{eq:rule_aa}
\end{equation}

which shows that the average link persistence is ${\tilde{\tau}}_{\rm n} = \langle z(t) \rangle / U$,
and that the $U$ links can be mapped to $\langle z(t) \rangle$ links with a persistence of unity.
\ref{eq:rule_a},~\ref{eq:rule_aa}, which are exact, provide a concise microscopic
description of entanglements. It can be understood as follows.

By observing the system over a time $T$, which covers a few times its longest relaxation time, 
we found that for the average single chain the entanglement environment is made by $U$ {\it different} 
chains, with which the constrained fluctuating PP collides and forms {\it binary links}. Each link
represents a pairwise uncrossability interaction, with a {\it specific} mate chain. The average linking
time, ${\tau}_{\rm n}$, is much smaller than $T$, because the majority of links are active only for very 
short times (blinking links). The large number of these interactions shows that the entanglement
environment is {\it inherently collective}. This description corresponds to the left-hand side of 
\ref{eq:rule_a}. 

The right-hand side is basically a mean-field type average of the left-hand side, where the
identities of the mate chains are mixed. 
It is understood as follows.
For the average single chain, the entanglement environment can be viewed as $\langle z(t) \rangle$
links per chain which are active over the entire observation time, $T$.
These links do not blink. Their linking time is $\tau = T$ (permanent links), so that their link 
persistence is ${\tilde{\tau}} = 1$. 
However, the identities of their mate chains, and therefore their coordinates in real space and
along the chain, are not specific. They are of mean-field type. 
Each one represents many pairwise interactions, and all together represent the complete 
set of the $U$ links on the left-hand side. 
This view leads to the foundations of the mean field picture employed in slip-link models. 

The above discussion is also valid in the limit $T \to \infty$. In this case, the appropriate
equation for the mapping of the collective entanglement environment to 
a small number of $\langle z(t) \rangle$ mean-field type interactions is \ref{eq:rule_aa}. 
In stochastic slip-link models, these mean-field interactions are usually {\it realized} as binary slip 
links, though they are not necessarily pairwise. For example, it is possible to consider that slip links 
represent multichain interactions, as has been discussed by Nair and Schieber~\cite{Nair-06}. 
The degree of multichain or pairwise character of these mean field interactions is 
examined in the next section. 



\subsection{Pairwise Parameter}
\label{section:PairPar}

\begin{figure} [!tbp]
\begin{center}
  \includegraphics[clip,width=0.8\linewidth] {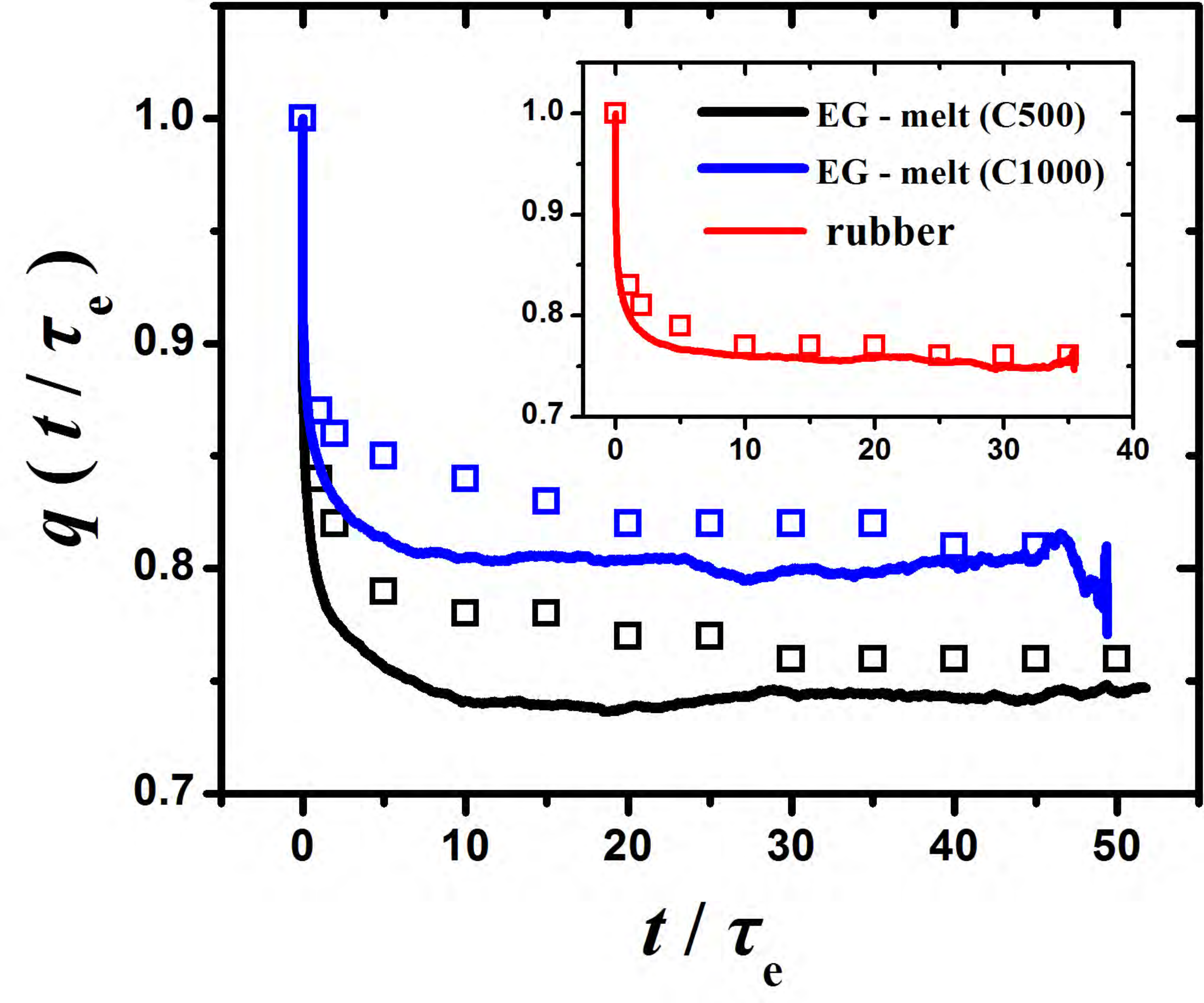}
\end{center}
\caption{\label{fig:q_t} 
  Fraction of links at time $t=0$, which are also present at time $t$.
  The plateau indicates the pairwise character of the mean-field interactions 
  that confine the average single chain.
  The squares denote corresponding predictions from the weight-average link persistence
  ${\tilde{\tau}}_{\rm w}(T)$, for observation times $(T / \tau_{\rm e})=0,~1,~2,~5,~10,~15,~...,~50$. (see text).}
\end{figure}

To examine the mean field uncrossability interactions in more detail, we define $z_{\alpha}(t)|z_{\alpha}(0)$ 
as the number of links at $t=0$, which also contribute to $z_{\alpha}(t)$ 
at time $t$, i.e., the `overlap' between the sets contributing to $z_{\alpha}(0)$ and $z_{\alpha}(t)$.
By definition, this number is smaller than or equal to $z_{\alpha}(0)$  (see~\ref{fig:zuq}). 
By averaging $z_{\alpha}(t)|z_{\alpha}(0)$ over multiple time origins along an equilibrium trajectory, 
we obtain the autocorrelation function $\langle z_{\alpha}(t)|z_{\alpha}(0) \rangle$.
Note that the links which contribute to $z_{\alpha}(t)|z_{\alpha}(0)$, for a different
time origin, need not always be the same. The ratio 

\begin{equation}
   q_{\alpha}(t) = \frac {\langle z_{\alpha}(t)|z_{\alpha}(0) \rangle} {\langle z_{\alpha}(t) \rangle} \label{eq:q_a}
\end{equation}

is the percentage of links at time $t=0$ which are also present at time $t$, irrespective of their 
state of entanglement with $\alpha$ at times between $0$ and $t$. 
The corresponding average over all chains, $q(t)$, is plotted in~\ref{fig:q_t}. It is
defined as 

\begin{align}
   q\left( t \right)&=\frac{\sum\limits_{\alpha =1}^{{{N}_{\rm{ch}}}}{\left\langle {{z}_{\alpha }}\left( t \right)|{{z}_{\alpha }}\left( 0 \right) \right\rangle }}{\sum\limits_{\alpha =1}^{{{N}_{\rm{ch}}}}{\left\langle {{z}_{\alpha }}\left( t \right) \right\rangle }}= \nonumber\\   
   \nonumber\\
  &=\frac{\left\langle z\left( t \right)|z\left( 0 \right) \right\rangle }{\left\langle z\left( t \right) \right\rangle } \label{eq:q}
\end{align}

We observe that $q(t)$ starts from unity ($t=0$), the largest possible value,
then decreases, and after a certain time reaches a plateau. For all systems, this time is around 
$t = 10 \tau_{\rm{e}}$. The decay of $q(t)$ is due to the mean-field nature of the 
$\langle z(t) \rangle$ links. The pairwise character of these links is quantified by the 
{\it pairwise parameter}, $q$, which is defined as the asymptotic value $q=q(t \to \infty)$. 
We find that $q$ is $0.76$ for the rubber, $0.74$ for the EG-melt (C500), and $0.81$ for the 
EG-melt (C1000). 

In principle, $0 < q \leq 1$, though it is obvious that it cannot become arbitrarily
  \bibnote{Note that, as explained in Appendix VII, the pairwise parameter can only be defined in a 
           system where all entanglements are trapped.}
small. According to \ref{eq:q}, when $q=0$, the chains linked to a reference chain at times zero
and $t$ would all be different to each other, something impossible in a system with fixed topology.
$q=1$ corresponds to a strictly {\it pairwise character}. The larger the deviation of $q$ from unity, the 
higher the multichain or {\it collective character} of the environment. Here, the mean field links have a 
strong pairwise character, supporting the employment of binary links in slip-link models. Moreover, it seems 
that the pairwise parameter is practically chain length and entanglement density independent, a very interesting 
result.

A simple model which relates $q$ to the weight-average link persistence,
and therefore to the confinement strength of pairwise interactions, is presented below.
The linking probability, $P_{\alpha \beta}$,
that chain $\alpha$ is linked to chain $\beta$, can be estimated as the ratio of the linking time 
${\tau}_{\alpha \beta}$, over the observation time $T$, i.e., $P_{\alpha \beta} = {\tilde{\tau}}_{\rm \alpha \beta}$. 
The probability that these chains are not linked is 
$1-P_{\alpha \beta}$. Obviously, $P_{\alpha \beta}=0$ for chains not included in $U_\alpha$.
The analysis below concerns long observation times, $T \to \infty$, 
where these probabilities become $T$-independent,with the convention $t \equiv T$.

The average number of different chains linked to chain $\alpha$  is

\begin{equation}
   \left\langle {{z}_{\alpha }}\left( t \right) \right\rangle =\sum\limits_{\beta =1}^{{{U}_{\alpha }}}{{{P}_{\alpha \beta }}}= \sum\limits_{\beta =1}^{U_{\alpha}} {{\tilde{\tau }}_{\alpha \beta}} \label{eq:Da}
\end{equation}

On the same footing, it is meaningful to assume that the time correlation function defined 
in \ref{eq:q_a} can be written as

\begin{equation}
   \left\langle {{z}_{\alpha }}\left( t \right)|{{z}_{\alpha }}\left( 0 \right) \right\rangle =\sum\limits_{\beta =1}^{{{U}_{\alpha }}}{{{P}_{\alpha \beta }}}{{P}_{\alpha \beta }} \label{eq:DaDa}
\end{equation}

Then, the weight-average persistence of the links of chain $\alpha$ reads 

\begin{align}
   {{\tilde{\tau }}_{\rm{w},\alpha}} &= \sum\limits_{\beta =1}^{{{U}_{\alpha }}}{{{w}_{\alpha \beta }}}{{\tilde{\tau }}_{\alpha \beta }}= \nonumber\\
   \nonumber\\
   &=\frac{1}{\left\langle {{z}_{\alpha }}\left( t \right) \right\rangle }\sum\limits_{\beta =1}^{{{U}_{\alpha }}}{{{P}_{\alpha \beta }}}{{P}_{\alpha \beta }} \label{eq:tau_w}
\end{align}

which leads to 

\begin{equation}
   \left\langle {{z}_{\alpha }}\left( t \right) \right\rangle {{\tilde{\tau }}_{\rm{w},\alpha}}=\left\langle {{z}_{\alpha }}\left( t \right)|{{z}_{\alpha }}\left( 0 \right) \right\rangle ~. \label{eq:rule_b1}
\end{equation}

As in \ref{eq:rule_a1}, 
$\langle z_{\alpha}(t) \rangle$ and ${\tilde{\tau}}_{{\rm w},\alpha}$ are independently 
distributed variables over the chains of the system (numerically checked).
Thus, the chain average of \ref{eq:rule_b1} reads 
\begin{equation}
   \left\langle z\left( t \right) \right\rangle {{\tilde{\tau }}_{w}}=\left\langle z\left( t \right)|z\left( 0 \right) \right\rangle \cdot 1  \label{eq:rule_b}
\end{equation}
From~\ref{eq:q},\ref{eq:rule_b}, and by taking into account that $T \to \infty$
we then find
\begin{equation}
    q = {\tilde{\tau}}_{\rm w}~~. \label{eq:qw}
\end{equation}
Moreover, \ref{eq:rule_b} can now be written as
\begin{equation}
    q U {\tilde{\tau}}_{\rm n} = \left\langle z\left( t \right) \right\rangle {{\tilde{\tau }}_{w}} ~,  \label{eq:rule_bb}
\end{equation}
which is the `weighted' form of \ref{eq:rule_aa}. 

A key result of this work is \ref{eq:rule_b}.
In the latter, the $\langle z(t) \rangle$ mean-field links, of strength-weighted average
link persistence ${\tilde{\tau}}_{\rm w}$, are mapped to 
$\langle z(t)|z(0) \rangle$ (or $q \langle z(t) \rangle$) links with a
strength-weighted persistence of unity (shown explicitly in the equation). 

\ref{eq:qw} is in very good agreement with the data of~\ref{fig:q_t}.
Even for short observation times $T$, where ${\tilde{\tau}}_{\rm w}$
is $T$-dependent (see Supporting Information), the relation $q(T) = {\tilde{\tau}}_{\rm w} (T)$ is able to follow the data,
while in the asymptotic regime the agreement is very good. 
The small deviations are due to \ref{eq:DaDa}, which holds only approximately.

The above eqs when ${\tilde{\tau}}_{\rm n} = {\tilde{\tau}}_{\rm w} = 1$, i.e., when
the links do not `blink', lead to $\langle z(t) \rangle = \langle {z(t)|z(0)} \rangle = U$, 
and $q = 1$. This is the case of a corresponding slip-link model, i.e., of a strictly pairwise entanglement environment
(see \ref{section:Distr}). In practical terms, the meaning of the pairwise parameter, and of \ref{eq:q}, 
is that the mean-field links can be realized as $q \langle z(t) \rangle$ {\it effective binary links},
with specific mate chain identities and coordinates.

%
%
%
%
%

\subsection{Selecting the Strongest Pairwise Interactions}
\label{section:Selection}

In \ref{section:Distr}, we have seen that the links have various strengths, 
which were categorized according to their link persistence,
$\tilde{\tau}$. However, a specific persistence scale separating the links into strong and weak does 
not seem to exist. The reason is that, for intermittent links, it is not only the linking time which 
qualifies a link, but also how this time is distributed over the observation time. The more 
uniformly it is distributed, the less weak is the link, as long as its persistence is appreciable.
Therefore, a methodology for selecting the strongest links has to consider the idle time intervals, 
or time gaps, $\tau_{\rm g}$, of the blinking links. These are the gaps in the `link trajectories' 
of~\ref{fig:U_a}.

\begin{figure} [!tbp]
\begin{center}
  \includegraphics[clip,width=0.8\linewidth] {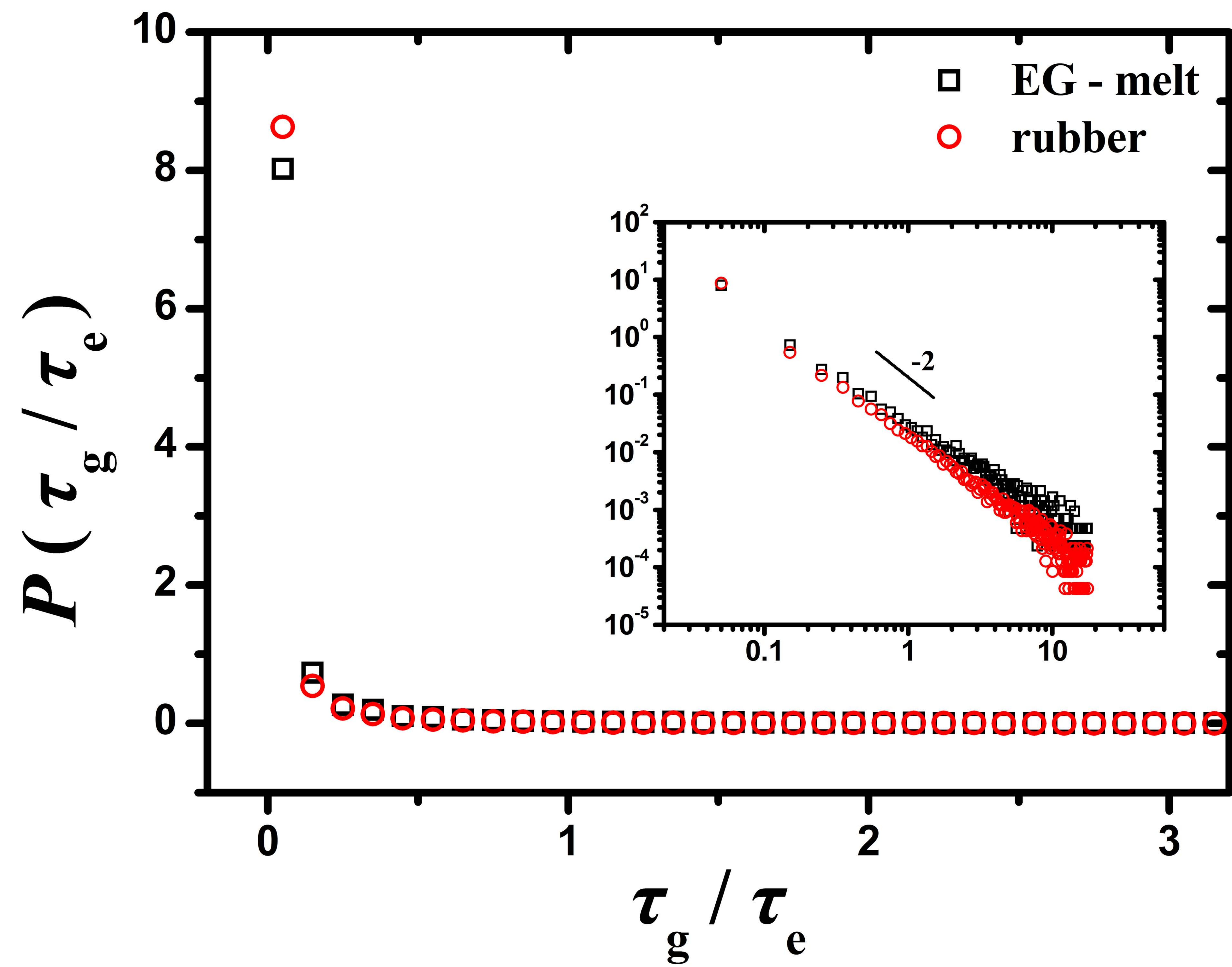} 
\end{center}
\caption{\label{fig:P_tau_g}
 Distribution of idle time intervals (time gaps), $\tau_{\rm g}$, of intermittent link `trajectories' such 
 as those presented in~\ref{fig:U_a}. Log-log coordinates in the inset.}
\end{figure}

In~\ref{fig:P_tau_g} we present the normalized distribution of $\tau_{\rm g}$ reduced
by $\tau_{\rm e}$.  The time gaps before the first, or after the last appearance of a
link contribute to the distribution. We see that it follows
a power-law decay (inset), and thus it is 
scale-free. The exponent is somewhat higher than -2. Almost all values are concentrated 
at $\tau_{\rm g} \ll \tau_{\rm e}$, and large time gaps are rare.

In order to select the strongest links, a possible strategy is to consider that, 
for idle time intervals less than or equal to a maximum $\tau_{\rm g,max}$, the links are 
instead active. This way we actually `fill in' certain time gaps
of intermittent link trajectories, such as those in~\ref{fig:U_a}.
When {\it all the time gaps} of a trajectory are less than or equal to the
specified $\tau_{\rm g,max}$ then this link spans the whole observation time, and it
can be selected as a link which is sampled at least once within this time interval. 
The idle times at the beginning and end of the observation time, if present, can be treated with a 
different $\tau_{\rm g,max}$. Here, for simplicity, a common $\tau_{\rm g,max}$ has been used. 

\begin{figure} [!tbp]
\begin{center}
  \includegraphics[clip,width=0.8\linewidth] {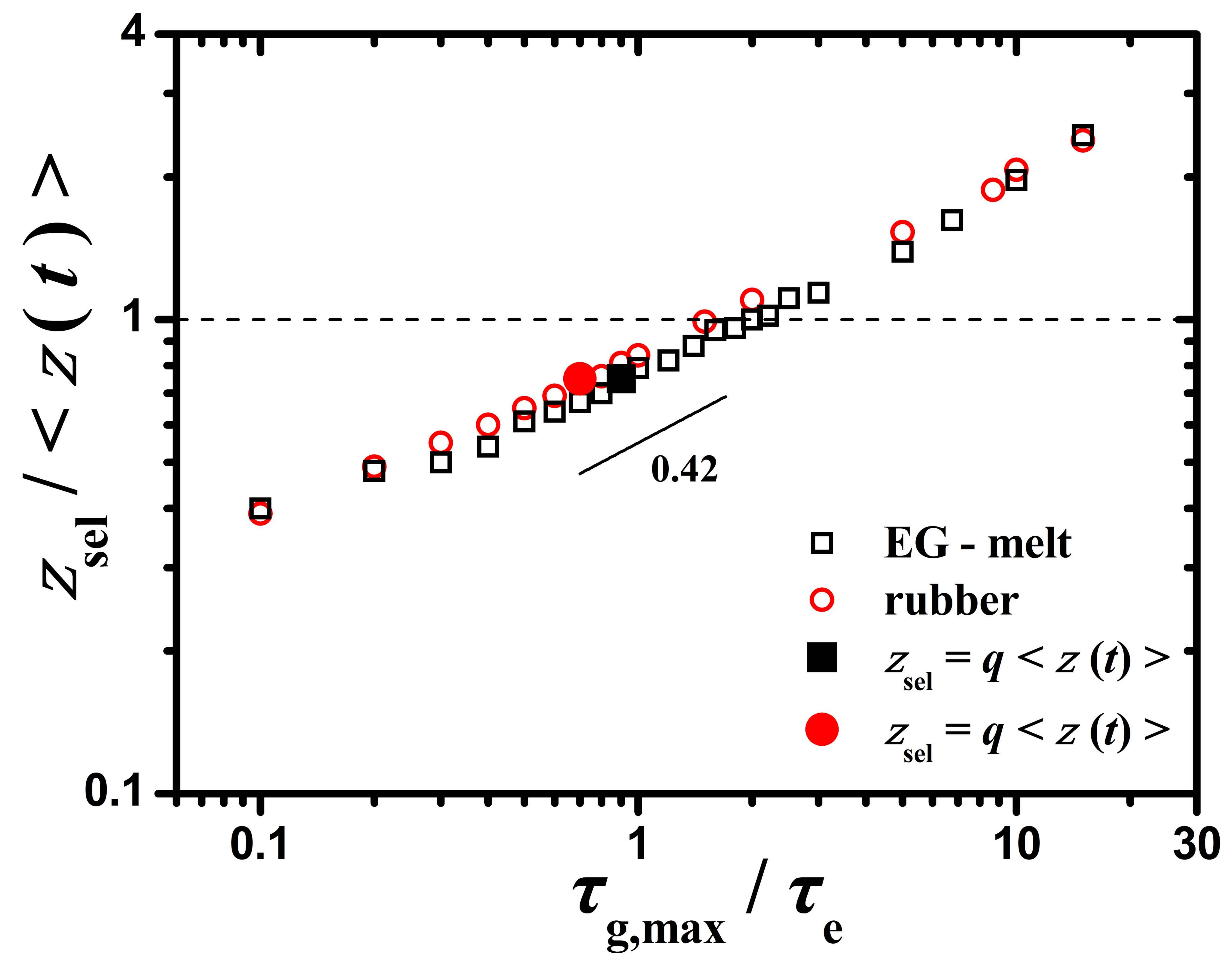}
\end{center}
\caption{\label{fig:D_g} 
 Ratio between selected links, $z_{\rm sel}$, and average number of links per chain 
 $\langle z(t) \rangle$ for various time gaps $\tau_{\rm g,max}$. 
 Filled symbols correspond to ratios dictated by the pairwise parameter, $q$.
 }
\end{figure}

The ratio of the average number of selected links per chain, 
$z_{\rm sel}$,  over $\langle z(t) \rangle$, for various values of $\tau_{\rm g,max}$, 
is shown in~\ref{fig:D_g}. In the regime of interest the ratio scales as 
$(\tau_{\rm g,max} / \tau_{\rm e})^{0.42}$. From this plot we can pick out an appropriate time scale 
for filling the gaps of link trajectories. Below we examine two possible choices, and
a third one is examined in the next section.

Following the outcome of weighted averaging, $\tau_{\rm g,max}$ can be adjusted so that 
we select as many links as the {\it effective binary links}, i.e., 
$z_{\rm sel} = q{\langle z(t) \rangle}$, where $q$ is the pairwise parameter. 
In~\ref{fig:D_g}, this case is shown with filled symbols, and the corresponding 
$\tau_{\rm g,max}$ is approximately one $\tau_{\rm e}$, for both systems. 

The other choice is to select as many links as the average number of links per chain, 
which are mean-field type links (see \ref{section:BinCol}). That is,
$z_{\rm sel} = {\langle z(t) \rangle}$. This scenario corresponds to a {\it `realization' of the
mean field} in terms of specific pairwise interactions and link coordinates. 
In~\ref{fig:D_g}, the corresponding $\tau_{\rm g,max}$ is estimated from the intersections of the dashed line 
with the data. It is $1.5\tau_{\rm e}$ for the rubber and $2\tau_{\rm e}$ for the EG-melt (C500).

\begin{figure} [!tbp]
\begin{center}
  \includegraphics[clip,width=0.8\linewidth] {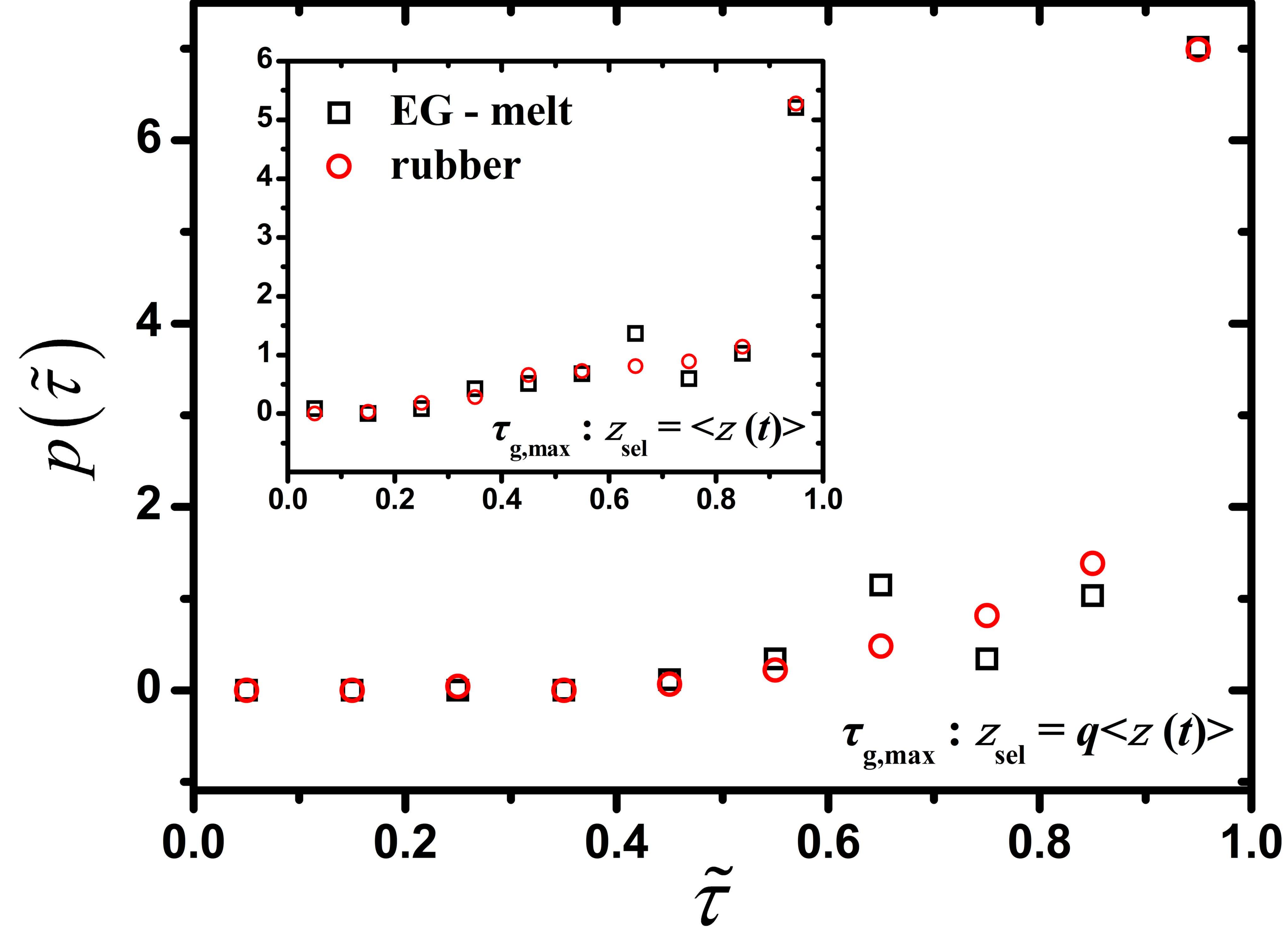}
\end{center}
\caption{\label{fig:p_g} 
 Distribution of the persistence of selected links.
 The maximum time gap, $\tau_{\rm g,max}$, is adjusted so that $z_{\rm sel} = q \langle z(t) \rangle$, 
 (filled symbols in~\ref{fig:D_g}). 
 In the inset the maximum time gap is adjusted so
 that $z_{\rm sel} = \langle z(t) \rangle$. The complementary distribution of discarded links can be found
 in the Supporting Information.}
\end{figure}

The time scale of the maximum time gaps is very reasonable. 
The distribution of the persistence of selected links is shown in~\ref{fig:p_g}.
The peak at weak links has now disappeared. We also see that a smaller
$\tau_{\rm g,max}$ leads to fewer selected links, with larger persistencies.
It also seems that the distributions are superposable. 
By adjusting $\tau_{\rm g,max}$ to fit the effective binary links, 
contributions from links with $\tilde{\tau} < 0.45$ vanish.  

\begin{figure} [!tbp]
\begin{center}
  \includegraphics[clip,width=0.8\linewidth] {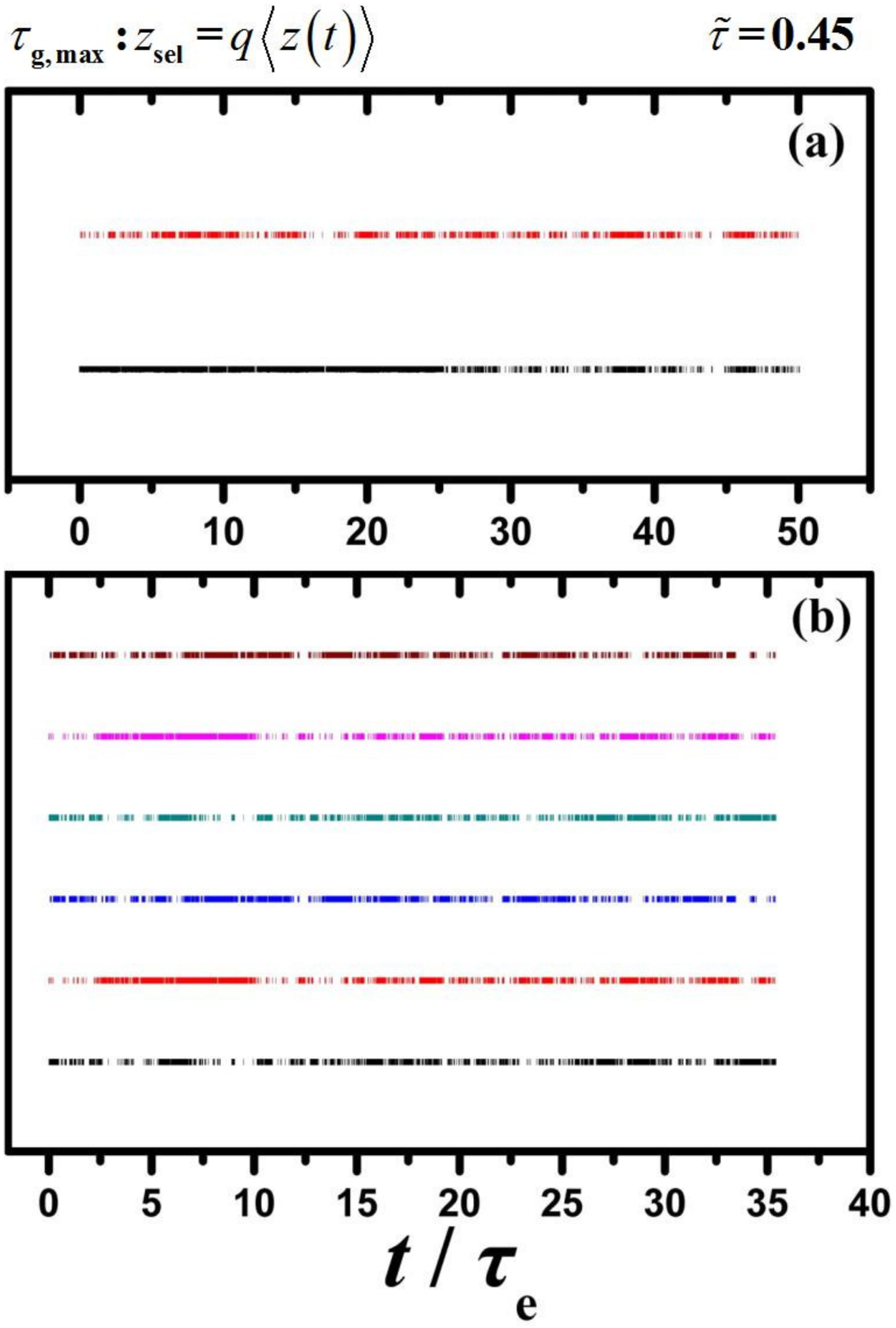}
\end{center}
\caption{\label{fig:links_g} 
  Intermittent link trajectories that correspond to {\it effective binary links}, discussed
  in \ref{section:PairPar}.
  They have been selected by using a maximum time gap 
  such that $z_{\rm sel} = q \langle z(t) \rangle$. (a) EG-melt (C500), (b) rubber.
  All the trajectories have a link persistence $\tilde{\tau} \simeq 0.45$.
  $\tau_{\rm g,max}$ is $0.7 \tau_{\rm e}$ for the rubber and one $\tau_{\rm e}$ for the EG-melt (C500).
  The trajectories are plotted with different colors  and parallel to each other.
  The vertical axis does not provide any kind of information.
}
\end{figure}

To examine the effectiveness of this strategy, in~\ref{fig:links_g} we plot
the trajectories of {\it all the links} selected under the condition 
$z_{\rm sel} = q\langle z(t) \rangle$,  with  $\tilde{\tau} \simeq 0.45$. 
There are few of them. It is seen that they blink in such a way that they cover the 
whole observation time. 

The above results show that the entanglement environment can be mapped 
to specific pairwise interactions, according to a given link density.
The selection procedure sorts out the 
links according to their strength, and then selects interactions from the strong part of the confinement 
strength spectrum. A certain time scale (interval), $\tau_{\rm g,max}$, is introduced, which is 
{\it self-consistently} determined to be of the order of one $\tau_{\rm e}$.  The selected links have the 
property that they are sampled at least once within this time interval. 

The use of this time scale permits a {\it controlled study} with
respect to a slip-link mapping of the system, and the development
of simulation schemes which track the motion of entanglements at equilibrium, 
or under deformation and flow. For trapped entanglements, this is a straightforward procedure. 
First, the strong links can be selected by analyzing a short equilibrium MD trajectory. 
Then, it is possible \cite{Stef} to follow these links upon any kind of deformation. 
The whole method is relatively fast, as long as the necessary software exists. 
When CR is active, an additional scheme \cite{Stef} for dealing 
with the update of links would be required. 

\subsection{Link Density, Storage Modulus and Slip-Link Mapping }
\label{section:Plateau}

Rheological estimates of $M_{\rm e}$ from 
plateau modulus measurements are in very good agreement with 
$M_{\rm e}$ estimates~\cite{Ever-04,Leon-05,Tzoum-06a,Tzoum-07,Kaz-07,Harm-09,Fotein-06}
from the Kuhn segment of the PPs, as obtained from atomistic systems.  
However, this length scale is larger \cite{Tzoum-06a,Tzoum-06b} than the mesh length,
\bibnote{In ref~\citenum{Tzoum-06a} $M_{\rm TC}$ is denoted as $M_{\rm ES}$.} 
$M_{\rm TC}$, of the networks presented here, 
i.e., $M_{\rm e} \simeq 2$ to $2.5 M_{\rm TC}$,
leading to a link density which is larger~\cite{Ever-12} than the `rheological' entanglement density. 

In the language of this paper, $M_{\rm TC}$ is the average molar mass between the successive, 
$\langle z(t) \rangle$ on average, links per chain. The statistical properties of these networks have
been presented elsewhere \cite{Tzoum-06a}. Conformationally speaking, the larger link density is 
due to exponentially decaying orientational correlations \cite{Tzoum-06a} along successive steps of 
the PP. A slowing down of Rouse modes conforming to the mesh length $M_{\rm TC}$ of the networks has
also been observed \cite{Tzoum-09b,Padd-02}. 

Hence, a one-to-one mapping of our networks to slip-link models employing the rheological entanglement
density, is not viable. However, given that the links here have various strengths, a way to resolve this
issue is to consider the PP length as obtained from the networks, but
for the slip-link motion under deformation, flow, or in equilibrium, 
to consider only a subset of very strong links. For example, in Appendix II we present the mean square 
displacement of the selected links $z_{\rm sel} = q {\langle z(t) \rangle}$, 
which was constructed by averaging over their individual real space trajectories. 

Then, a practical approach is to adjust $\tau_{\rm g,max}$ of the previous section, so that the 
selected link density matches the rheological entanglement density. Applying this scheme here, 
we found that only the very strong links, with a persistence $\tilde{\tau} > 0.96$, were selected (see below). 
The required $\tau_{\rm g,max}$ is very small, approximately $0.1 \tau_{\rm e}$, for both systems. 
It remains to be shown that the spatial statistics of these links provide a meaningful mesoscopic picture. 
This approach seems promising as a simple coarse graining scheme for bridging the gap between atomistic 
and slip-link models. We note that this scheme is self-consistent, since $\tau_{\rm g,max}$ is adjusted by 
the $M_{\rm e}$ obtained from the Kuhn length of PPs. Alternatively, $\tau_{\rm g,max}$ can be adjusted so 
that the selected link density matches a relaxation modulus obtained from simulation.

Evidently, the most important question is how the trapped links revealed here contribute to the
low frequency storage modulus of an elastomer. 
For example, we have seen that the majority of the links blink instead of being continuously present (\ref{section:Cumul}).
The blinking is due to stored length fluctuations (Appendix V), which at the network level lead to thermal PP fluctuations and 
considerable idle times in certain link trajectories (\ref{fig:U_a}). Stored length fluctuations can `relax' the
constraining effect of these links, either fully or partially. Most probably, weak links can be fully relaxed so that they 
do not contribute to the modulus, while strong links can only partially relax. 
In this respect, the persistence of each link is indicative of its contribution to the low frequency modulus.

\begin{figure} [!tbp]
\begin{center}
  \includegraphics[clip,width=1.0\linewidth] {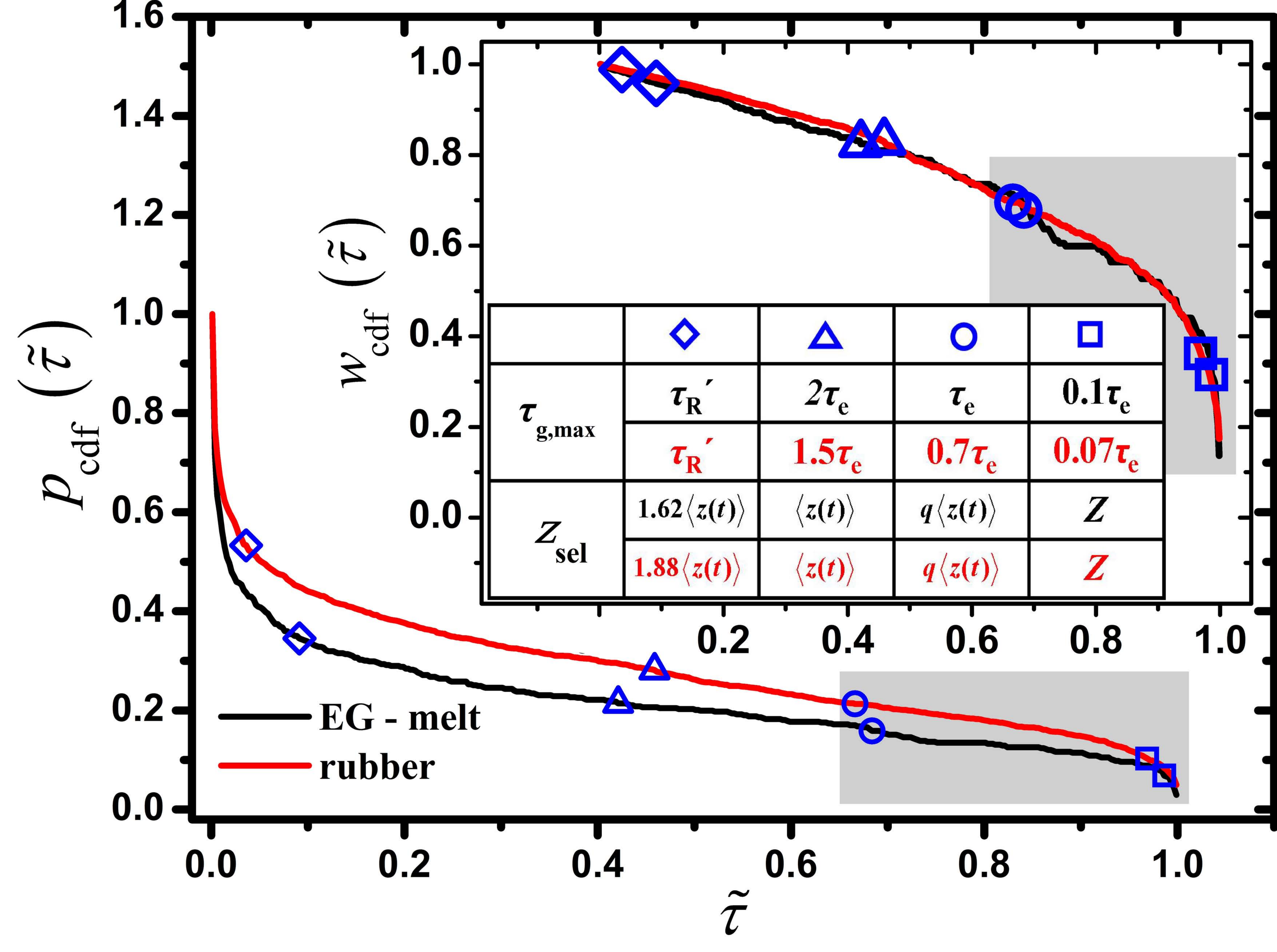}
\end{center}
\caption{\label{fig:pw_cdf} 
  Cumulative distribution function of $p \left(\tilde{\tau}\right)$, and $w \left(\tilde{\tau}\right)$ (inset), 
  which were presented in~\ref{fig:pw}. The symbols are put at the specific $\tilde{\tau}$'s for which 
  $U p_{\rm cdf} \left(\tilde{\tau}\right) = z_{\rm sel}$, according to the table of the inset (see text). The curves
  should be examined from the right to the left. The shaded area encloses the spectrum of $\tilde{\tau}$'s 
  corresponding to links which can potentially contribute to the low frequency modulus of an elastomer, or to
  the plateau modulus of a melt. 
}
\end{figure}

To investigate this issue further, we define the cumulative distribution function (CDF) of the persistence of links
\begin{equation}
       p_{\rm cdf} \left(\tilde{\tau}\right) = \int_{\tilde{\tau}}^1 \! p\left({\tilde{\tau}}'\right) \, \mathrm{d} {\tilde{\tau}}'~~,   \label{eq:cdf}
\end{equation}
where $p\left(\tilde{\tau}\right)$ is integrated backward, from the right to the left. 
$p_{\rm cdf} \left(\tilde{\tau}\right)$ is the fraction of $U$ links with a persistence $\tilde{\tau} \le {\tilde{\tau}}' \le 1$.
We also introduce $w_{\rm cdf} \left(\tilde{\tau}\right)$, the CDF corresponding to $w \left(\tilde{\tau}\right)$, 
which is defined similarly to $p_{\rm cdf} \left(\tilde{\tau}\right)$. 
$w_{\rm cdf} \left(\tilde{\tau}\right)$ is the fraction of confinement strength provided by links with a persistence 
$\tilde{\tau} \le {\tilde{\tau}}' \le 1$. Both distributions are shown in~\ref{fig:pw_cdf}. 
From the form of the CDFs we see that a $\tilde{\tau}$-scale separating the links into strong and weak is not obvious. 
However, we observe that the two $w_{\rm cdf} \left(\tilde{\tau}\right)$ superpose, and that there are two regimes.
From $\tilde{\tau} \simeq 1$ to $\tilde{\tau} \simeq 0.95$ the confinement strength provided by these links increases abruptly, 
while below $\tilde{\tau} \simeq 0.95$ it follows a smooth increase. 

A way to examine~\ref{fig:pw_cdf} with respect to relaxation effects is the following.
In the inset we present a table with the required maximum time gap, $\tau_{\rm g,max}$,
so that the number of selected links, $z_{\rm sel}$, matches certain link densities discussed in the article. 
$Z = \left( {\langle L_{\rm pp} \rangle} / d  \right) - 1$, where $\langle L_{\rm pp} \rangle$ 
is the average PP contour length, and $d$ is the PP Kuhn length obtained by mapping the PPs to random walks~\cite{Tzoum-06a}
(see Appendix II). $Z$ corresponds to the number of `rheological' entanglements, i.e., $Z = (M / M_{\rm e}) - 1$,
and is much smaller than $\langle z(t) \rangle$. 

It is meaningful to assume that links with a large $\tau_{\rm g,max}$ can relax to a larger degree than links with 
a small $\tau_{\rm g,max}$. For example, we can safely consider that the links with $\tau_{\rm g,max} \geq {\tau'}_{\rm R}$, where 
${\tau'}_{\rm R}$ is the longest monomer relaxation time in the system (Appendix II), 
do not contribute to the low frequency modulus. 
On the other hand, links with $\tau_{\rm g,max} \le \tau_{\rm e}$ can contribute to the modulus, but they
can also partially relax, possibly to a different degree. Grossly speaking, we can assume that at low frequencies 
links with $\tau_{\rm g,max} > \tau_{\rm e}$ become fully relaxed. We envisage that links which disappear and
reappear, with idle times longer than $\tau_{\rm e}$, have been relaxed by stored length fluctuations.  

To proceed further we need to associate a pair of $\left( z_{\rm sel}, \tau_{\rm g,max} \right)$, to a link persistence. 
This can be done by associating $z_{\rm sel}$ to a specific $\tilde{\tau}$,
such that the fraction of the $U$ links with $\tilde{\tau} \le {\tilde{\tau}}' \le 1$ is $z_{\rm sel}$. 
That is, $U p_{\rm cdf} \left(\tilde{\tau}\right) = z_{\rm sel}$. For example, each symbol of~\ref{fig:pw_cdf}  
is plotted at a $\tilde{\tau}$ associated with a specific pair $\left( z_{\rm sel}, \tau_{\rm g,max} \right)$ of the table in
the inset. 
We see that the fraction of links with $\tau_{\rm g,max} < \tau_{\rm e}$ is approximately 20\%, with 
$\tilde{\tau} > 0.65$, for both systems (circle symbols). These links provide approximately 70\% of the confinement strength, (see $w_{\rm cdf} \left(\tilde{\tau}\right)$)
they are $q \langle z(t) \rangle$ in number, and they can potentially contribute to the low frequency modulus. 
In~\ref{fig:pw_cdf} they are enclosed in a gray shaded area. 

The rheological entanglement density corresponds to $Z$ very strong links with
$\tau_{\rm g,max} \simeq 0.1\tau_{\rm e}$ and $\tilde{\tau} > 0.96$ (square symbols). These links provide only
30\% to 40\% of the confinement strength. Interestingly, the full confinement strength is provided by
links with $\tau_{\rm g,max} < {{\tau}'}_{\rm R}$ (diamond symbols).
The above results demonstrate that weak links, though essentially trapped, do not contribute to the low 
frequency modulus, in contrast to common belief. 
It  seems that stored length fluctuations can fully relax weak topological interactions.
The role of these fluctuations in the viscoelasticity~\cite{Edwa-00} of elastomers and melts is central.
Regarding the contribution of weak links to the plateau modulus of a melt, 
the above discussion is still valid.


\subsection{Discussion}
\label{section:Discussion}

We have seen that the pairwise uncrossability constraints detected as kinks by chain shrinking methods, 
such as the Z1~\cite{Krog-05,Fotein-06}, CReTA~\cite{Tzoum-06a} algorithms, 
or the primitive path analysis method \cite{Ever-04,Leon-10}, can be either strong or weak interactions,
and not detected at all times (link blinking), even in the absence of CR. 
The entanglement environment is collective, and certain binary interactions are more important than others.
Thus, under nonlinear deformation or flow, where some kind of entanglement network evolution takes place 
(due to CR), it is possible that changes in the number of detected kinks \cite{Leon-10,Maha-10,Toepp-11}
do not reflect the actual microscopic changes of the entangled state of the system. 

\section{Summary and Conclusions}
\label{section:Conclusions}

In molecular theories of rubber elasticity and polymer rheology,
the microscopic entanglement constraints are replaced with a confining mean field, 
that restricts laterally chain dynamics and fluctuations.
Here, we examine in detail at what level this mean field is a collective entanglement
effect, as in tube model theories, or can be described through certain pairwise 
uncrossability interactions, as in slip-link models (see p. 552 in ref \citenum{Graess-08}).

To this end, for the average single chain, the entanglement environment was expanded to a set of
$U$ {\it local links}. Each link represents a {\it pairwise} uncrossability interaction with a 
{\it different} neighboring chain, resembling a familiar slip-link.
Since in our systems all entanglements
are {\it trapped}, we claim that the $U$ pairwise interactions form a {\it complete} set, and that 
the entangled state of the system, at any time, can be projected to them. Even if some 
interactions have been left out of the analysis, it is shown that they are too weak to be considered.

The number of the $U$ interactions, or links, is very large, which means that the entanglement environment 
is inherently collective. However, their confinement strength varies from strong to weak,
and the majority of them are weak. The plethora of these links is due to the fact that Primitive Paths 
(PPs) are fluctuating objects. Thus, as a PP fluctuates, it collides with the contours of other PPs
and forms local links with them. This sampling mechanism reveals that weak links appear as intermittent, 
blinking links, because they are not continuously sampled. However, there also exist strong links which 
are sampled continuously, and appear as permanent constraints. The weak links can "disappear" for time 
intervals longer than $\tau_{\rm e}$, thus they do not contribute to the low frequency modulus of an 
elastomer, or the plateau modulus of a melt. Overall, the average effect of these
interactions is to confine a chain through a much smaller number of mean-field type, permanent links, $\langle z(t) \rangle$,
which in principle represent many other pairwise interactions.

The multichain, or pairwise character of these mean-field links was quantified by
defining a pairwise parameter, and it was shown that the pairwise character dominates.
Moreover, for the systems studied the pairwise parameter
is entanglement density and chain length independent. These results imply that  
the realization of a confining mean field with binary interactions, as in slip-link models, 
is not in contrast with a collective entanglement environment. 

On this basis, we proposed a self-consistent scheme for mapping the entanglement environment 
to a slip-link model. The scheme discards weak links and selects the strongest binary interactions
conforming to a given entanglement density. It is directly applicable to trapped entanglements,
and to the analysis of MD deformation trajectories of bead-spring and atomistic polymer networks.
From such studies, available slip-link network models can be refined, corrected, 
or extended, in a hierarchical manner. With suitable extensions \cite{Stef}, the tracking of 
constraint release events in polymer melt simulations, seems also possible. A list of symbols and
abbreviations for this paper is given in Table 1.

\section*{Appendix I: Construction of a Defect-Free Entangled PE Network}

In flexible polymers, the degree of chain overlap is controlled~\cite{Graes-81}
by density and stiffness. Therefore, it seems unusual that the two systems examined here (rubber and end-grafted melt)
have a different degree of entanglement, though the slightly larger $C_{\infty}$ of the rubber would decrease~\cite{Graes-81} 
$M_{\rm e}$, in comparison to the melt. Since both systems are well equilibrated,
the difference must be due to the short subchains and the artificial preparation conditions of the network. 
The latter is constructed as follows. At the nodes of the diamond lattice we place carbon atoms. 
The lattice constant is then adjusted so that the carbon atoms are bonded with PE chains at full extension 
(all-trans state of dihedral angles). This is an unentangled state. Then, the network is let to relax at 
1 Atm. Because of entropic elasticity and cohesive interactions it contracts to the melt density. At this stage the system is not 
equilibrated conformationally. To impose equilibration a phantom chain simulation at melt density 
(NVT ensemble) is performed. Bead overlaps are then eliminated by the gradual push-off~\cite{Auhl-03} 
method.  

Entanglements are introduced dynamically after the chains have been linked
to a network, by letting them interpenetrate at the phantom chain simulation stage. It seems that this
procedure, together with the short length of the chains, leads to a larger degree of chain overlap in 
comparison to the melt state. We expect that in networks of very long subchains, where the cross-link 
density vanishes, this effect would subside. It can also be avoided through random linking of a melt 
system, at the cost of introducing network defects. On the other hand, the fact that our results have been 
obtained with different entanglement densities makes our conclusions fairly general. 

A different path to defect-free networks has been followed by Everaers~\cite{Ever-96,Ever-99}.
It invokes constructing identical interpenetrating diamond networks, where the lattice constant is set to 
the root mean squared end-to-end distance of melt chains. The system consists of as many networks as are 
required to reach the melt density, which is held fixed. As in our case, equilibration is achieved by 
letting the networks interpenetrate at a phantom chain simulation stage. Chain overlaps are then eliminated 
through a gradual~\cite{Auhl-03} `push-off' stage. In comparison to our networks the chains can also 
entangle with chains from other identical networks. 

\section*{Appendix II: Mean Square Displacement and Characteristic Relaxation Times}

\begin{figure} [!tbp]
\begin{center}
  \includegraphics[clip,width=0.8\linewidth] {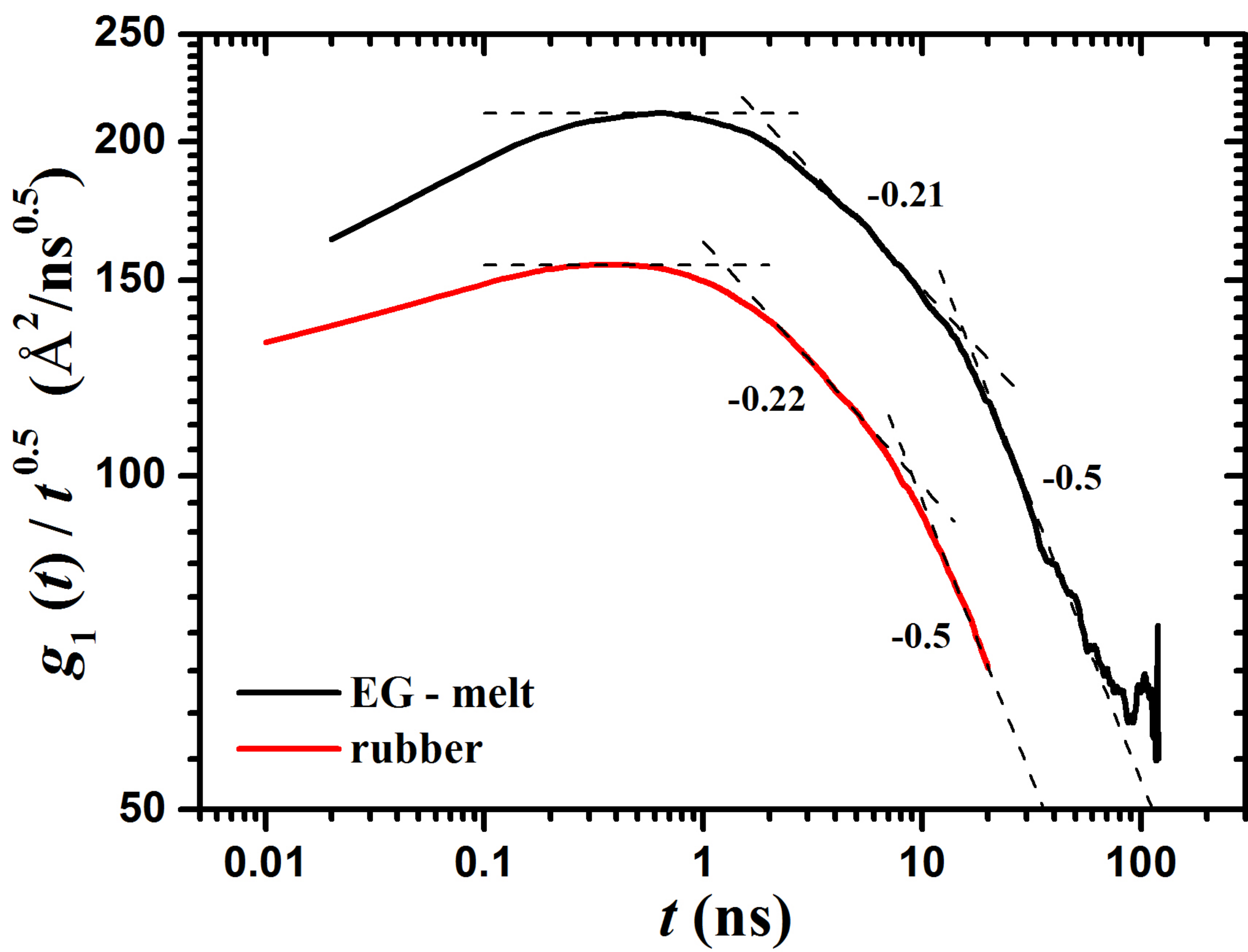}
\end{center}
\caption{\label{fig:msds} 
Mean square displacement of middle chain monomers (10\%), reduced by $t^{0.5}$, for the rubber and
the (C500) EG-melt.
From the crossovers we estimate the shortest and longest relaxation time of each system.}
\end{figure}

In~\ref{fig:msds} we present the mean square displacement (msd), $g_{1}(t)$, of middle chain monomers
(10\% of each chain), reduced by $t^{0.5}$. Four regimes are discernible.
The first one is unimportant. It is usually called ballistic. 
A second regime, scaling as $t^{0.5}$, corresponds to
free Rouse motion within the tube. The next regime is indicative of tube confinement (restricted Rouse 
motion along a random-walk like tube), and it scales, approximately, as $t^{0.28}$, for both systems.
The corresponding tube model prediction~\cite{deGen-79} is $t^{0.25}$. The last regime corresponds to an
invariant msd. From the first crossover (between second and third regime) we estimate
$\tau_{\rm e}$, the time that a monomer `hits' the tube for  the first time. It is 1.3 ns for the rubber 
and 1.7 ns for the EG-melt. From the second crossover we estimate ${\tau'}_{\rm R}$, the longest
monomer relaxation time of each system, 
which is 9.8 ns for the rubber and 11.5 ns for the EG-melt. This time corresponds to the
Rouse time, $\tau_{\rm R}$. Here, it is denoted with a prime, since the boundary conditions
for a Rouse chain are different (free chain ends).
For the (C1000) EG-melt, the corresponding crossover times are 1.6 and 19.4 ns (not shown).

\begin{figure} [!tbp]
\begin{center}
  \includegraphics[clip,width=0.8\linewidth] {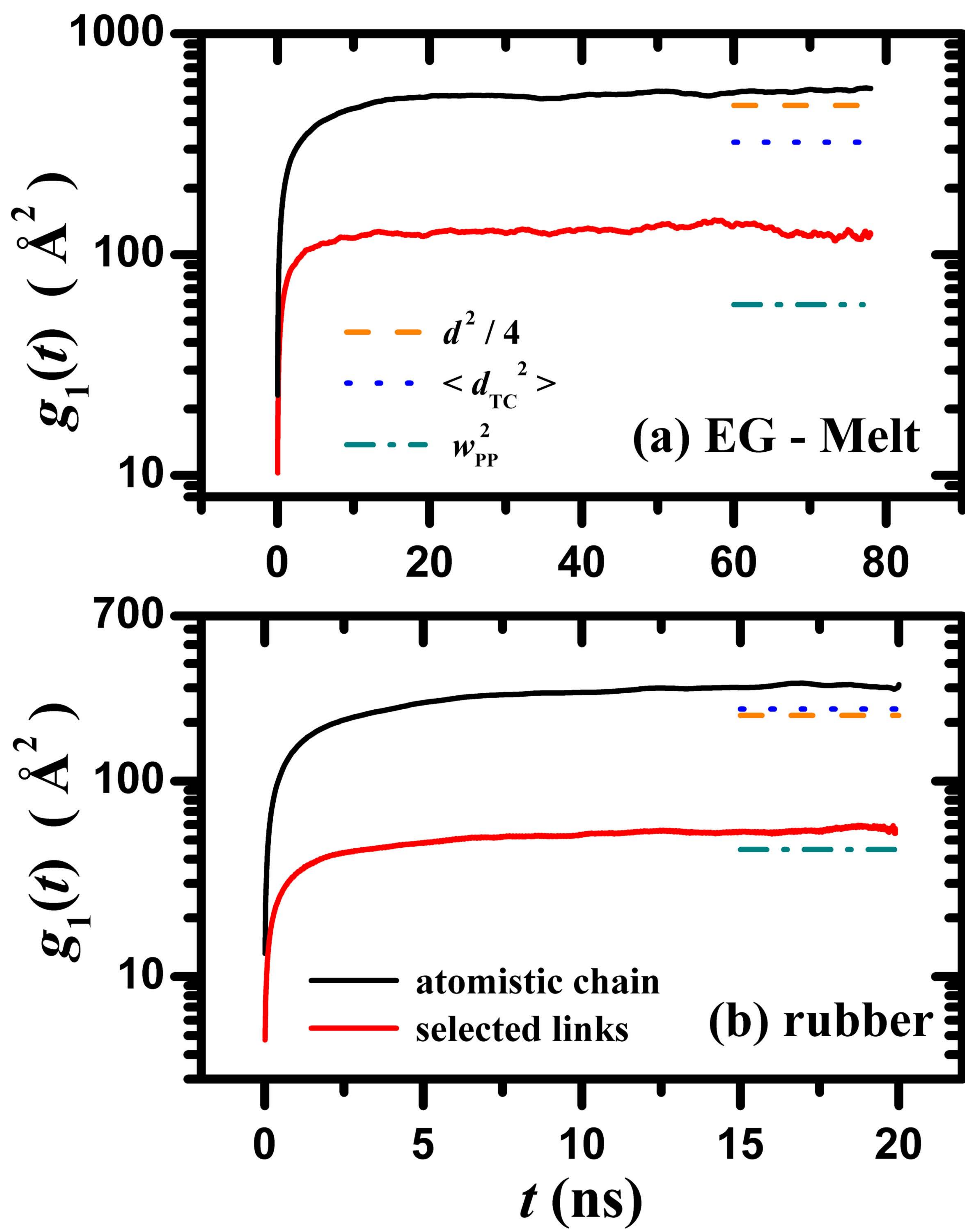}
\end{center}
\caption{\label{fig:msd_SL} 
Mean square displacement of middle chain monomers, and selected links. 
(a) EG-melt (C500); (b) rubber. 
The links were selected according to the condition $z_{\rm sel} = q \langle z(t) \rangle$.
The bars parallel to the plateaus show certain length scales related to entanglement 
constraints (see text).}
\end{figure}

In~\ref{fig:msd_SL} we present the msd of middle chain monomers
together with the msd of links selected according to the 
pairwise parameter, i.e., $z_{\rm sel} = q \langle z(t) \rangle$. 
The plateaus indicate the absence of reptation and long chain diffusion.
We see that the selected links span much shorter distances than real chain monomers.
Thus, we can say that the enclosing tube of PPs is a shrunk version
of the alleged tube that restricts lateral motion of real chain monomers.

The bars plotted parallel to the plateaus correspond to estimations of certain length scales 
related to entanglement constraints. $d$ is the tube diameter estimated from the Kuhn segment of 
the PP as $d = \langle R^2 \rangle/\langle L_{\rm pp} \rangle$, where $\langle R^2 \rangle$ is the mean square 
end-to-end distance. 
$\langle d_{\rm TC}^2 \rangle$ is the mean square edge length of the link network~\cite{Tzoum-06a,Tzoum-09a}. 
$w_{\rm pp}^2$ is an estimation of the mean square distance between PP contours~\cite{Tzoum-06a,Tzoum-09a}
from the inverse of the PP contour length density. 
It is defined as $w_{\rm pp}^2 = \langle V \rangle/(N_{\rm ch}\langle L_{\rm pp} \rangle)$, 
where $\langle V \rangle$ is the average volume. Note that the average distance between PP contours is smaller
than the average distance between connected links along the PPs, $\langle d_{\rm TC} \rangle$. For example,
in~\ref{fig:hooking}, the distance between PP contours, $h$, is much smaller than the PP segments
with average length $\langle d_{\rm TC} \rangle$.  

For both systems, we see that the msd plateau of the monomers shows a better correlation with $d^2 / 4$, 
which corresponds to the squared radius of the tube, but is actually larger than that. This is not
surprising, since real chain monomers can penetrate the tube walls, as parts of unentangled loops 
for example. In both systems we also observe a correlation between the squared tube
radius and the mean edge length of the link network. The msd plateaus of selected links are much 
smaller than these quantities, which means that link fluctuations are restricted to the inner tube region along the tube axis. 
These plateaus are also larger than $w_{\rm pp}^2$, especially in the EG-melt, which means that the 
fluctuating PPs can collide to each other, in accordance with the lateral link sampling mechanism.

\section*{Appendix III: Binary Links, Local Knots, and Ternary Nodes}

As shown in \ref{fig:links}, each node links locally two chains, except from few ternary nodes
(see below). Within this terminology, each node resembles a binary link that can be resolved into two TCs 
along two {\it{ `mate chains'}}. Iwata and Edwards have discussed~\cite{Iwa-88,Iwa-89} 
melt dynamics and rheology in terms of local knots which are conceptually similar to the ones presented here. 
The knots were defined at the real chain level by using as topological criterion a local form of the 
Gauss linking number. For open chains this criterion is not more rigorous~\cite{Tzoum-11,Mill-05,Orlan-04}
than the one defined here, while it is computationally more complex.

In our networks about 5\% of the nodes are ternary. 
That is, there exist situations where three chains, let us say $\alpha$, $\beta$, and $\gamma$,
are mutually constrained and form local links with each other (see videos). 
Detection of these cases is hard and needs special~\cite{Stef} attention.
They are mapped to nodes with six edges, which are further 
resolved~\cite{Stef} into three links, $\alpha\beta$, $\alpha\gamma$, $\beta\gamma$, 
so that all three chains are mutually coupled in pairs.
With this choice, the mapping to a network leads exclusively to binary links,
as in slip-link models. Information about ternary nodes is retrievable, though.

Upon dynamical evolution of the network, most of the ternary nodes tend to 
break up into consecutive binary nodes along the same chain, e.g., 
$\alpha\beta$, $\alpha\gamma$, in the former example. The reverse procedure is also possible (see videos).
A statistical analysis \cite{Tzoum-06a} has shown that between successive links of a chain there exists an effective repulsion. 
Thus, we expect that, while ternary links certainly appear, they are transient. Interestingly, a repulsive interaction
also emerges when enforcing~\cite{Uney-11} detailed balance in the dynamical equations of the slip-link based primitive 
chain network model. 

\section*{Appendix IV: Multiply Linked Chains}

\begin{figure} [!tbp]
\begin{center}
  \includegraphics[clip,width=0.8\linewidth] {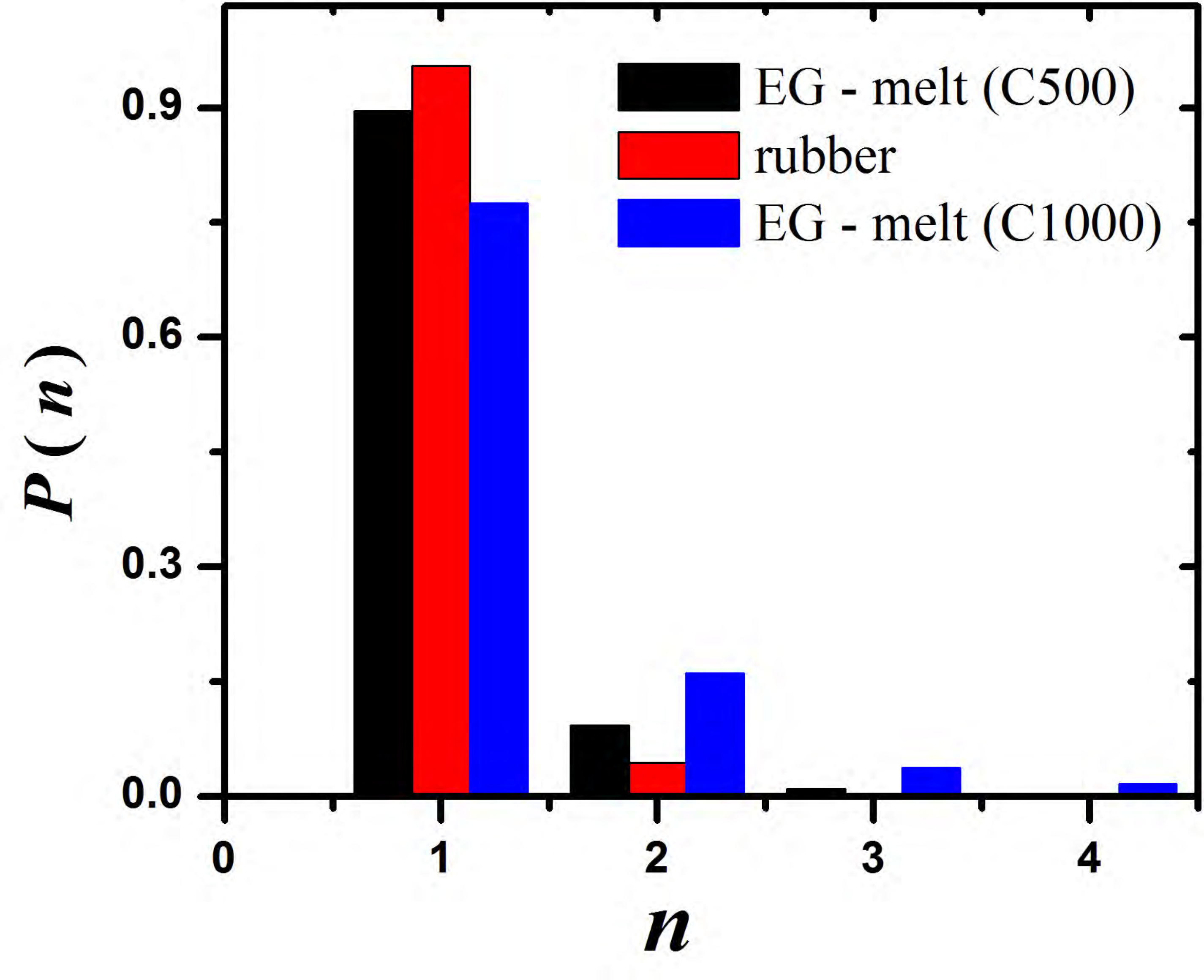}
\end{center}
\caption{\label{fig:p_n} 
    Normalized probability that a chain forms $n$ local links with another mate chain.}
\end{figure}

In~\ref{fig:p_n} we show the normalized probability, $P(n)$, that a chain $\alpha$ forms 
$n$ local links with a mate chain in $z_{\alpha}(t)$, instantaneously, at time $t$, (see also~\ref{fig:zuq}). 
For the rubber and the short chain EG-melt $P(n)$ is highly peaked at unity. 
Therefore, the simplification of dealing with LCs instead of 
individual links is justified. However, as chain length increases $P(1)$ decreases.
Thus, in systems with long chains the analysis should be made against individual links.

\section*{Appendix V: Lateral Link Sampling and CReTA Mapping}

The large set of different chains sampled by the PP is rather surprising.
Since the PPs are determined by contour reduction, then, by construction,
the sampled chains have portions inside or at the boundaries of the pervaded volume
of a reference chain. Moreover, they are not just simple 
chain contacts, since linked chains satisfy the local topological criterion of~\ref{fig:links},
(although some very weak links could be considered as fluctuating chain contacts).
The lateral link sampling (LLS) mechanism that takes place is simple and was demonstrated in~\ref{fig:hooking}. 

\begin{figure} [!tbp]
\begin{center}
  \includegraphics[clip,width=0.8\linewidth] {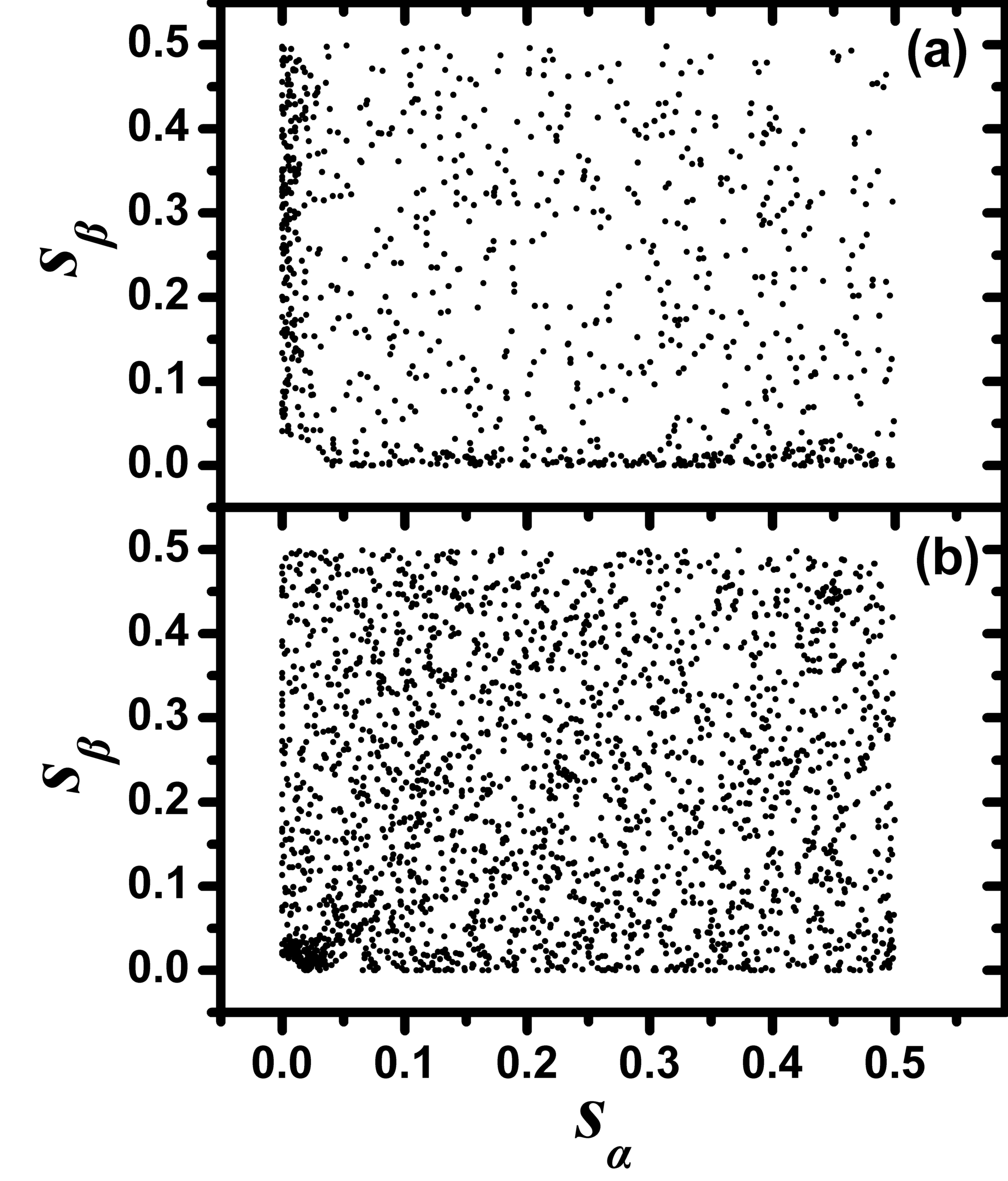}
\end{center}
\caption{\label{fig:s-s} 
Reduced curvilinear coordinates of the links sampled by the mechanism of~\ref{fig:hooking}:
(a) EG-melt (C500); (b) rubber. In order to increase statistical accuracy we exploit head to tail 
symmetry. Coordinates with $s>0.5$ are plotted at $|1-s|$. 
}
\end{figure}

In order to provide more evidence for this mechanism, we first define the 
reduced curvilinear coordinates of a link, $s_\alpha(t)$, $s_\beta(t)$ between 
chains $\alpha, \beta$. These are defined such that $0 \leq s \leq 1$.
That is, $s_{\alpha}(t)=x_{\alpha}(t)/L_{\alpha}(t)$, where $x_{\alpha}(t)$ is the
curvilinear coordinate of the link along chain $\alpha$, and $L_{\alpha}(t)$ is the PP contour length.
Likewise for $s_\beta(t)$. Note that a local link can slide along PP contours, 
hence the time dependence of $s_\alpha$, $s_\beta$.

In~\ref{fig:s-s} we plot $s_\alpha(t)$, $s_\beta(t)$, 
for {\it all the links} that were first sampled at a time $t>0$. In~\ref{fig:U_a}, these links
correspond to the trajectories with $U_{\rm{\alpha}}(t) > U_{\rm{\alpha}}(0)$, 
(continuous or segmented trajectories above the column to the left of $t=0$ in ~\ref{fig:U_a}). 
For each one of these links, we plot {\it a single point} that denotes $s_\alpha(t)$, $s_\beta(t)$,
at the time the link was first sampled.
Beyond some explainable heterogeneities, the homogeneous patterns support the LLS
mechanism of~\ref{fig:hooking}. The EG-melt data are less dense due to the smaller number
of simulated chains.

Data concentration at the lower left corner of~\ref{fig:s-s}b
implies that many of the new links are formed between chain ends. 
This is explained by the higher probability of instantaneous link formation
between PPs attached to the same tetrafunctional crosslink of a network.
In the EG-melt, local links with $s_{\rm{\alpha}} \simeq 0$ 
and $s_{\rm{\beta}} \simeq 0$ are favored.
This is an artificial effect, due to the less mobile chain ends because of grafting. 
The fluctuating PPs have higher probability to collide with certain PP segments which
are relatively immobile, or attached to fixed points.
In systems with very long chains these heterogeneites would disappear.

The physical origin of the LLS mechanism is the thermal motion of real chains and, more
specifically, the thermal motion of stored length~\cite{deGen-79}, or chain slack. 
The {\it spatial} distribution of this length is different at different time-snapshots
of our systems. In slip-link models this `redistribution' is taken into
account by the thermal motion of links and by a random transfer of Kuhn segments between neighboring 
strands of the PP. Similar mechanisms apply in our simulations.
Since CReTA operates by consuming chain slack,
the spatial redistribution of stored length leads to associated thermal fluctuations
of PP conformations (in real space), and transfer of monomers between the strands of the PP. 
As a result, there exist small displacements of network nodes which lead to the LLS mechanism.

As regards the mapping of a specific spatial distribution of stored length to a network of
links by CReTA, this is not unique. CReTA minimizes the total contour length of the system. 
Assuming that there is a unique, `global' minimum, the value of the objective 
function will hardly be changed if we remove some small length from a
PP and add it to another. The stochastic nature of the reduction performed by 
CReTA is subject to this effect.
At the network level, this will lead to small displacement of nodes, 
and therefore, possibly, to link sampling by the mechanism of~\ref{fig:hooking}. 

Thus, individual CReTA reductions of the same configuration (same 
spatial distribution of stored length), performed with different pseudorandom numbers,
may sample, each time, a slightly different set of links, with practically the same shortest 
paths (PP conformations and length). This uncertainty is inherent to the reduction problem. 
Equivalently, one could say that the stored 
length can only be mapped to a small irregular volume around the centerline of the corresponding tube 
(see~\ref{fig:tube}), and not to a specific path. This volume encloses a subset of the links which 
define the tube constraint and has large overlap with the tube core. 
An individual reduction by CReTA, then, will sample a path line in this volume. The links of the subset 
which are inside the tube core will be sampled with higher probability. This kind of 
inherent `noise' in the network mapping does not affect any of our conclusions. 
It is very small~\cite{Stef} relative to the thermal noise 
driving the sampling of the links that assemble the tube constraint.

\section*{Appendix VI: Entanglement Constraints at the Level of Shortest Paths and Local Links}

\begin{figure} [!tbp]
\begin{center}
  \includegraphics[clip,width=0.8\linewidth] {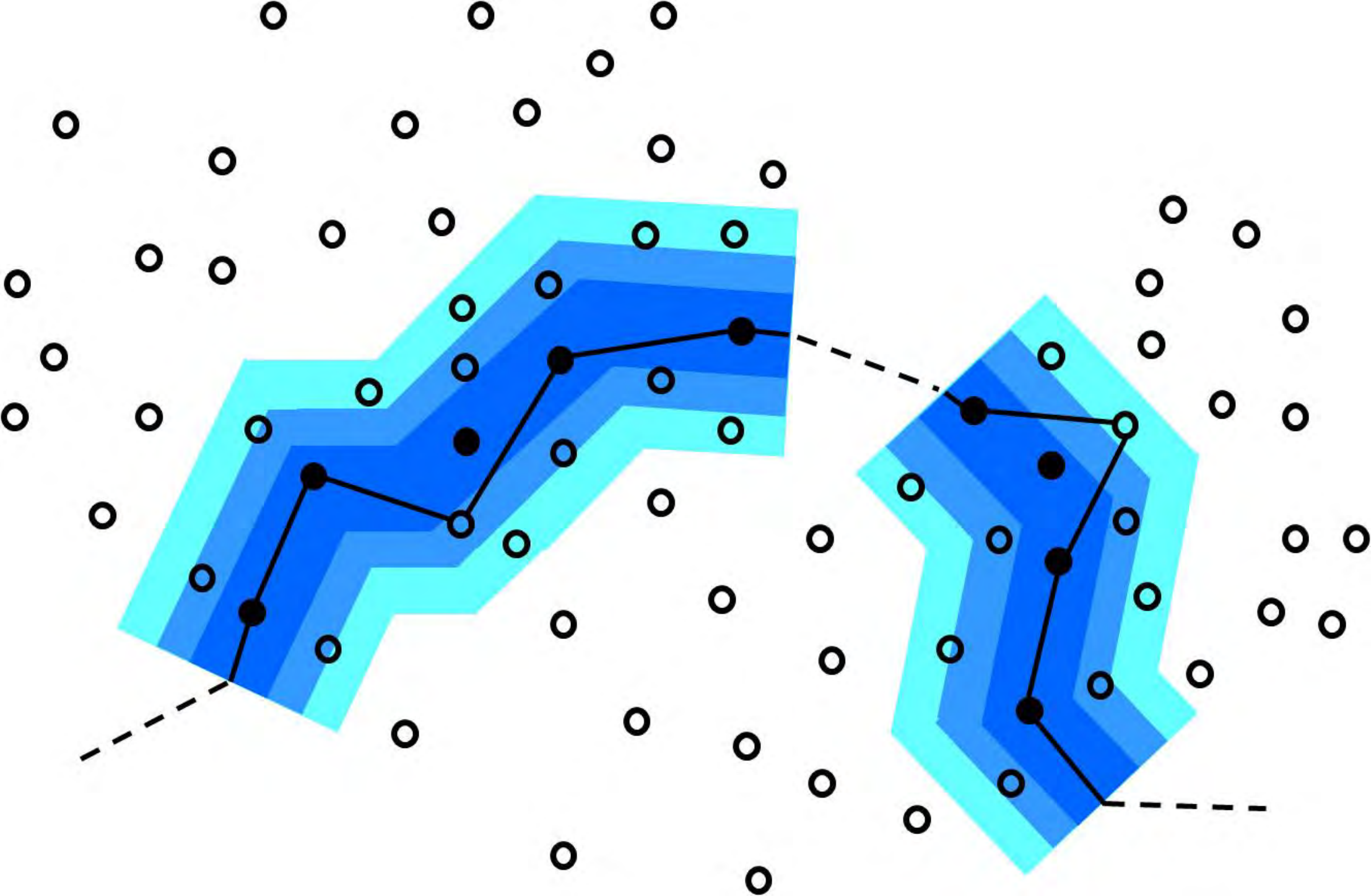}
\end{center}
\caption{\label{fig:tube}
Schematic representation of entanglement constraints, at the level of local links. 
A chain $\alpha$ in a system with fixed tubes is considered.
The PP is a fluctuating object outlined with a black line. The dashed lines denote parts of the 
PP which are not shown.
The open and filled circles denote the fluctuating PPs of other chains, which cross the
figure plane at various directions. Because of lateral fluctuations of
all PP conformations, the PP of $\alpha$ can collide with the other PPs denoted by circles,
and form local links with them.
Therefore, the PP of $\alpha$ samples its environment by tracing a fluctuating path through these links.
The sampling of links inside the shaded tube, which are $U_{\alpha}$ in number, 
does not violate uncrossability constraints.
Links outside the tube cannot be sampled, or are too weak to be considered.
After a certain time, which for the systems studied is $T \simeq 40-50 \tau_e$, 
(and depends on entanglement density), 
the PP has sampled all the $U_{\alpha}$ links within its tube.
Instantaneously, the PP is linked to $z_{\alpha}(t)$ links, a subset of $U_{\alpha}$. 
The linking time (LT)  of individual links, $\tau$, varies significantly. 
In the figure, the tube constraint is colored blue along the {\it{tube core}} and light blue on 
the exterior, conveying the fact that the inner links (filled circles)
are sampled more frequently than outer ones. Thus, there exist {\it strong} and {\it weak} links,
and the tube constraint is applied {\it collectively} by all of them. 
In the provided videos weak links appear as {\it{intermittent, blinking nodes}}. 
The strong links, making the tube core, appear as nonblinking, continuously present nodes.
The diameter of the depicted tube is much smaller for PP segments~\cite{Stef}, than for real chain monomers 
(Appendix II), which are not restricted by the shaded tube. 
The mean distance between the circles is proportional to the average distance between PP contours.
Note that this distance is smaller than the 
average mesh length of the link network. The former distance is based on the density of sampled links 
$U_{\alpha}$, which is larger than the link density defined by 
$\langle z_{\alpha}(t) \rangle$ (\ref{section:Cumul}). Effectively, this means that the average distance between 
the circles is smaller than the average distance between the filled
circles along the tube axis. For example, in~\ref{fig:hooking}, the distance between PP contours, $h$, 
is much smaller than the PP segments with average length $<d_{\rm TC}>$, which on the average make
the tube axis. The sampling of links is described in~\ref{fig:hooking}, and in Appendix IV. 
}
\end{figure}

A full view of what happens at the link level requires the 
knowledge of the distances spanned by fluctuating
PPs. A short account is given in Appendix II. A more detailed presentation will be given elsewhere~\cite{Stef}. 
For the systems studied here, it suffices to say that, as time proceeds, the PP segments cover an increasing 
distance (through lateral fluctuations), until they have fully sampled their corresponding tubes.
Because of the coarse graining procedure they span shorter distances than the real chain 
monomers. Thereby, their enclosing tubes are shrunk versions of the real chain tubes,
and while they restrict lateral PP fluctuations they do not restrict lateral monomer
fluctuations. We could say that a PP samples the domain around the centerline of the real
chain tube. The picture that emerges is described in~\ref{fig:tube}.

\section*{Appendix VII: Constraint Release Effects on $q(t)$}

\begin{figure} [!tbp]
\begin{center}
  \includegraphics[clip,width=0.8\linewidth] {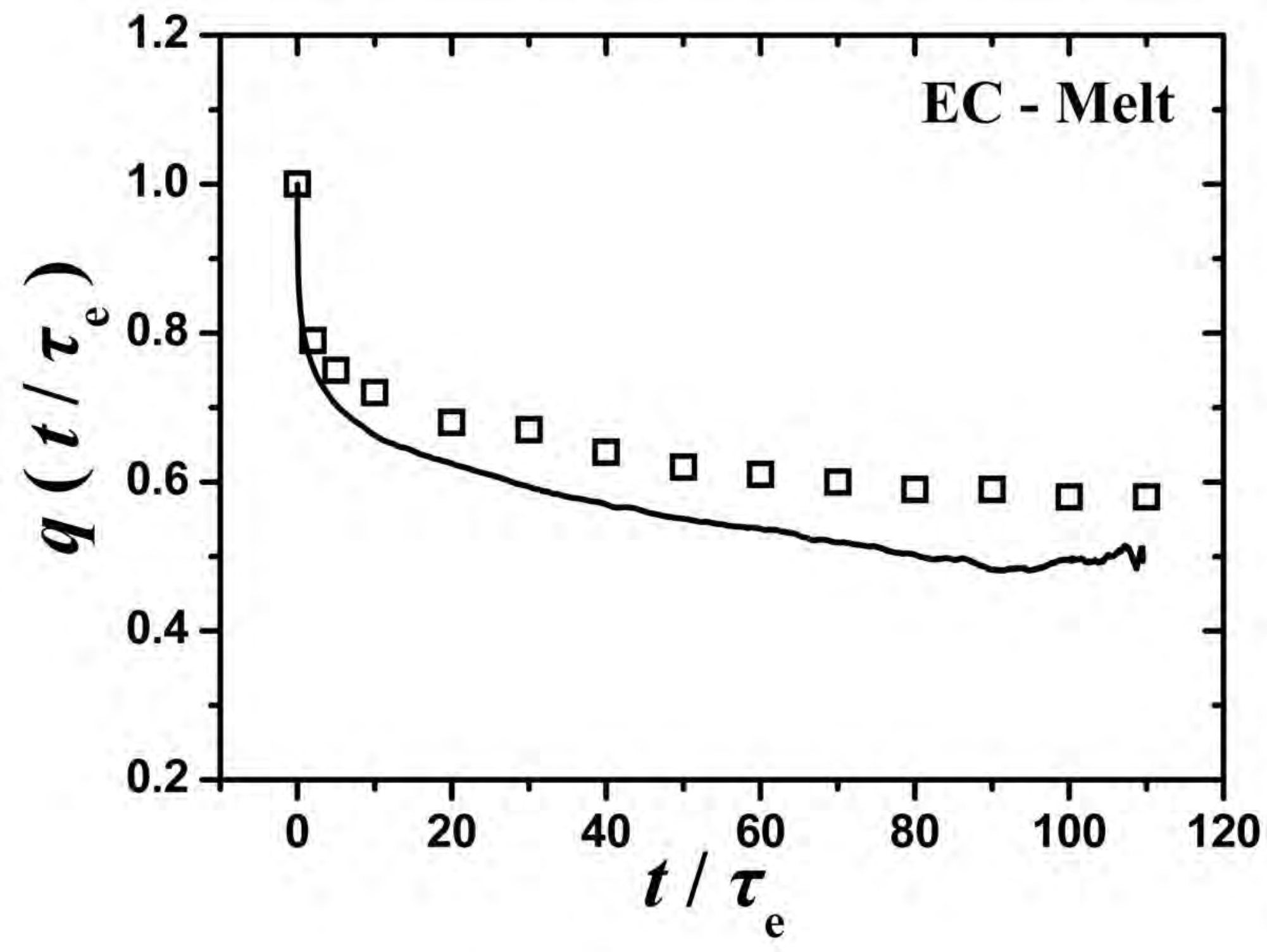}
\end{center}
\caption{\label{fig:q_ec} 
 Fraction of links at time $t=0$, which are also present at time $t$, for the
 end-constrained melt (C500) generated from the 
 same melt configuration as the EG-melt (C500).
 The squares denote $q(t)$ estimations from 
 the weight-average link persistence, as described in \ref{section:PairPar}.
}
\end{figure}

In~\ref{fig:q_ec} we present $q(t)$ for the (C500) melt, but this time analyzed as an end-constrained 
melt (EC-melt). That is, instead of grafting chain ends to nearby chains, the MD trajectory was generated 
by keeping all chain ends fixed. This difference has a strong effect on $q(t)$ due to the presence of a 
CR mechanism, which can be called end-looping constraint release (EL-CR). A polymer chain, say chain A, 
can disentangle from a mate chain, say chain B, by developing a lateral unentangled hernia that 
passes over a chain end of B. That is, by hopping of an internal portion of A over one end of B, 
through Rouse motion (see Figure 7 in the work of Zhou~\cite{Zhou-06} and Larson). 
Here, we see that EL-CR reduces $q(t)$ to 
a lower plateau value, which becomes discernible at $t / \tau_{\rm e} \simeq 100$. 
However, a safe estimation of this plateau would require a longer observation time.

Grossly speaking, the EC-melt can be viewed as the EG-melt with some links not being trapped, but subject 
to updating by EL-CR. This is similar to a network with dangling chains, i.e., with both trapped and 
temporary links. In such a system $q(t)$ will decay toward a certain plateau, characteristic of the
density of trapped entanglements. In a melt with free chain 
ends (exclusively temporary links), $q(t)$ will eventually vanish. 
The plateau value for the EC-melt will be smaller than that of the EG-melt (smaller trapped 
entanglement density). However, the pairwise parameter is defined as the plateau of $q(t)$
when a system is examined with all of its entanglements being trapped (fixed tubes). For this purpose, 
as explained in \ref{section:Method}, CR is eliminated by grafting all free chain ends.
Therefore, a plateau in $q(t)$, in the presence of CR, does not correspond to a pairwise parameter
for the entanglement environment. The decay of $q(t)$, or the difference between the 
EG-melt and EC-melt plateau, can possibly provide some kind of CR information, 
but this issue was not examined here.

\subsection*{Author Information}
{\bf{Corresponding Author}}

E-mail: tzoum@central.ntua.gr

\subsection*{\small Supporting Information}
  
  Entanglement network views of a PE melt (Figure S1), 
  link trajectories for representative chains (Figure S2),
  number and weight distribution of link persistences, for increasing observation time $T$ (Figure S3),
  Number-average and weight-average link persistence as a function of increasing $T$ (Figure S4),
  distribution of unselected link persistences (Figure S5), magnified section of~\ref{fig:pw_cdf} (Figure S6), and 
  15 explanatory videos for the material discussed in the document, with an associated description.  
  This material is available free of charge via the Internet at http://pubs.acs.org. 

\newpage

\begin{table}[!tbp] 
\caption{\label{table1}~List of Symbols and Abbreviations}
\begin{tabular}{ll}  
\hline 

$N_{\rm ch}$                            & Number of chains in the system   \\
$\alpha$                                & Chain index, $1 \le \alpha \le N_{ch}$    \\
$\tau_{\rm e}$                          & entanglement time \\
$\tau_{\rm R}$                          & Rouse time   \\
$M_{\rm e}$                             & Entanglement Molecular Weight  \\
$T$                                     & Observation Time (OT)  \\
$\tau$                                  & Linking Time (LT)  \\
$\tilde{\tau} = \tau / T$               & Link persistence \\
$z_{\alpha}(t)$                         & Number of different links or Linked Chains (LCs) \\
                                        & constraining $\alpha$ at time $t$ \\
$U_{\alpha}(t)$                         & Cumulative number of the different links \\
                                        & sampled by $\alpha$ up to time $t$ \\
$z_{\alpha}(t)|z_{\alpha}(0)$           & Subset of links constraining $\alpha$ at both times, $0$ and $t$ \\
$\langle z_{\alpha}(t)|z_{\alpha}(0) \rangle$     
                                        & $z_{\alpha}(t)|z_{\alpha}(0)$ averaged over multiple time origins\\
$q_{\alpha}(t)$                         & ratio of $\langle z_{\alpha}(t)|z_{\alpha}(0) \rangle$ over $\langle z_{\alpha}(t) \rangle$ \\
$z(t), U(t), q(t)$                      & Chain averages of the corresponding quantities above. \\
$q$                                     & Pairwise parameter, $q = \displaystyle \lim_{t \to \infty}q(t)$ \\
$p(\tilde{\tau})$                       & Number distribution of link persistence \\
$w(\tilde{\tau})$                       & Weight distribution of link persistence \\
${\tilde{\tau}}_{{\rm n},\alpha}$       & Average persistence of links of chain $\alpha$ \\
${\tilde{\tau}}_{\rm n}$                & Average link persistence \\
${\tilde{\tau}}_{{\rm w},\alpha}$       & Weight average persistence of links of chain $\alpha$ \\
${\tilde{\tau}}_{\rm w}$                & Weight average link persistence \\
${\tilde{\tau}}_{\alpha \beta}$         & Persistence of a link between chains $\alpha$, $\beta$ \\
$w_{\alpha \beta} ({\tilde{\tau}}_{\alpha \beta})$  & Confinement strength of a link with persistence ${\tilde{\tau}}_{\alpha \beta}$ \\
$P_{\alpha \beta}$                      & Probability that chain $\alpha$ is linked to chain $\beta$ \\
$\tau_{\rm g,max}$                      & Maximum time gap of link trajectories \\
$z_{\rm sel}$                           & Number of selected links per chain\\
PP                                      & Primitive Path \\
TC                                      & Topological Constraint \\
LC                                      & Linked Chain \\
LLS                                     & Lateral Link Sampling \\
CDF                                     & Cumulative Distribution function \\
EG-melt                                 & End-Grafted melt \\
EC-melt                                 & End-Constrained melt \\
EL-CR                                   & End-Looping Constraint-Release \\

\end{tabular} 
\end{table} 

\newpage
\newpage
\newpage
\newpage

\begin{acknowledgement}
  The authors would like to acknowledge collaborative work with Mark Robbins,
  Ting Ge, and Robert Hoy on the problem of quantifying entanglement loss in crazed polymer
  glasses, which helped in posing some of the problems examined here.  We also would like to thank 
  Athanasios Morozinis for sharing the MD trajectories used in this study.
  The work of C.T. and D.N.T., in the context of this paper, is
  part of the Research Programme of the Dutch Polymer Institute (DPI),
  Eindhoven, The Netherlands, Project No. 650. S.D. Anogiannakis
  is thankful to the National Technical University of Athens for a PhD fellowship.
  The authors declare no competing financial interest.
\end{acknowledgement}


\newpage

\bibliography{bibpart1}

\begin{mcitethebibliography}{124}
\bibitem[Not()]{Note-1}
Extreme care should be taken when dealing with different chain images. For
  readers not familiar with simulations, we note that the chain parts which are
  inside the simulation box can belong to different chain images. Here, 76 and
  329 different chains pass through the simulation box of the EG-melt (C500)
  and rubber system, respectively. The results are not influenced by periodic
  boundary conditions.\relax
\mciteBstWouldAddEndPunctfalse
\mciteSetBstMidEndSepPunct{\mcitedefaultmidpunct}
{}{\mcitedefaultseppunct}\relax
\EndOfBibitem
\bibitem[Not()]{Note-2}
By estimating the ratio of chain length over $(\langle z(t) \rangle + 1)$,
  which is proportional to the entanglement molar mass, we can compare the link
  density between the two systems. In the rubber it is about twice that of the
  EC-melt. Estimating $M_{\rm e}$ from the Kuhn segment of the PPs we are led
  to the same conclusion. Note that the link density of our systems is larger
  than the rheological entanglement density, as discussed
  in~\ref{section:Discussion}.\relax
\mciteBstWouldAddEndPunctfalse
\mciteSetBstMidEndSepPunct{\mcitedefaultmidpunct}
{}{\mcitedefaultseppunct}\relax
\EndOfBibitem
\bibitem[Not()]{Note-3}
On this point, strong evidence is provided by $q(t)$, introduced in Section
  5.5, which reaches a plateau at times where $U(t)$ is still increasing.\relax
\mciteBstWouldAddEndPunctfalse
\mciteSetBstMidEndSepPunct{\mcitedefaultmidpunct}
{}{\mcitedefaultseppunct}\relax
\EndOfBibitem
\bibitem[Not()]{Note-5}
Note that, as explained in Appendix VII, the pairwise parameter can only be
  defined in a system where all entanglements are trapped.\relax
\mciteBstWouldAddEndPunctfalse
\mciteSetBstMidEndSepPunct{\mcitedefaultmidpunct}
{}{\mcitedefaultseppunct}\relax
\EndOfBibitem
\end{mcitethebibliography}

\end{document}